\documentclass[fleqn,usenatbib]{mnras}

\usepackage{newtxtext,newtxmath}

\usepackage[T1]{fontenc}
\usepackage{ae,aecompl}

\usepackage{graphicx}	
\usepackage{amsmath}	

\title[Stacking 21cm Images Around High Redshift Galaxies]{Stacking Redshifted 21cm Images of HII Regions Around High Redshift Galaxies as a Probe of Early Reionization}

\author[James E. Davies]{
James E. Davies,$^{1,2}$\thanks{E-mail: daviesje@student.unimelb.edu.au}
Rupert A. Croft,$^{1,2,3}$
Tiziana Di-Matteo,$^{1,2,3}$
\newauthor
Bradley Greig,$^{1,2}$
Yu Feng,$^{4}$
and J. Stuart B. Wyithe$^{1,2}$
\\
$^{1}$School of Physics, The University of Melbourne, Parkville, Victoria 3010, Australia\\
$^{2}$ARC Centre of Excellence for All Sky Astrophysics in 3 Dimensions (ASTRO 3D)\\
$^{3}$McWilliams Center for Cosmology, Carnegie Mellon University, Pittsburgh PA, 15213\\
$^{4}$Berkeley Center for Cosmological Physics, University of California at Berkeley, Berkeley, CA, 94720, USA
}

\date{Accepted XXX. Received YYY; in original form ZZZ}

\pubyear{2019}

\begin{document}
\label{firstpage}
\pagerange{\pageref{firstpage}--\pageref{lastpage}}
\maketitle

\begin{abstract}
A number of current and future experiments aim to detect the reionization of neutral hydrogen by the first stars and galaxies in the Universe via the redshifted 21cm line. Using the \textsc{BlueTides} simulation, we investigate the measurement of an \textit{average} ionised region towards the beginning of reionization by stacking redshifted 21cm images around optically identified bright galaxies using mock observations. We find that with an SKA 1000 hour observation, assuming perfect foreground subtraction, a $5\sigma$ detection of a stacked HII region can be made with 30 images around some of the brightest galaxies in \textsc{bluetides} (brighter than $M_{UV} < -22.75$) at $z=9$ (corresponding to a neutral fraction of 90.1 \% in our model). We present simulated relationships between the UV magnitude of galaxies, the sizes of the ionised regions they reside in, and the shape of the stacked profiles. These mock observations can also distinguish between scenarios where the IGM is in net emission or absorption of 21cm photons. Once 21cm foreground contamination is included, we find that even with up to 200 images around these rare, bright galaxies, only a tentative $> 1\sigma$ detection will be possible. However, partial foreground subtraction substantially improves signal-to-noise. For example, we predict that reducing the area of Fourier space dominated by foregrounds by 50 (80) percent will allow $> 3\sigma$ ($> 5\sigma$) detections of ionised regions at $z=9$.
\end{abstract}




\section{Introduction}
The reionization of cosmic hydrogen is believed to have originated in the densest regions in the Universe, where bright star-forming galaxies were more highly clustered. These first ionised regions are predicted to grow and overlap, filling the IGM by $z \sim 6$ (e.g: \cite{Greig17}).


Radio observations of the redshifted 21cm line from the hyperfine splitting of atomic hydrogen currently present the most promising future probe of the epoch of reionization. Radio interferometers aim to detect the EoR in one of two ways. Most first generation low frequency radio telescopes such as the Murchison Widefield Array (MWA)\footnote{http://www.mwatelescope.org/} and the Low Frequency Array (LOFAR)\footnote{http://www.lofar.org/}, along with the more sensitive second generation telescope Hydrogen Epoch of Reionization Array (HERA)\footnote{https://reionization.org/} aim to detect the epoch of reionization statistically by measuring the 21cm power spectrum. Another second generation telescope, the Square Kilometre Array (SKA)\footnote{https://www.skatelescope.org/}, also aims to image reionization directly in 21cm emission \citep{Mellema13}. In this paper we focus on images of ionised regions around bright galaxies at high redshift. A number of works have considered identifying HII regions around individual sources using matched filters and imaging \citep{Datta07,Datta08,Majumdar12,Ghara17,Ghara19,Hassan19}, and similar methods using high-redshift quasars were proposed in \citet{Wyithe05,Kohler05,Geil08,Ma20}. Since these methods work with single images, they require relatively large bubbles, and nearly perfect subtraction of radio foregrounds in order to detect the EoR. Thus, imaging an individual HII region toward the beginning of reionization will be difficult. However, since galaxies are much more common than quasars, we can stack 21cm images around selected bright galaxies, in analogy to high-z HI detections \citep{Blyth15}, taking advantage of synergies between 21cm and optical-infrared observations in order to average out noise and boost the signal, facilitating observations of smaller HII regions.

Simulations suggest that the brightest galaxies at high redshift will likely sit in the first ionised regions. Not only do these galaxies produce a large number of ionising photons, but they are also more likely to be inside the dense, clustered filaments of the cosmic web, with many neighboring galaxies and quasars which also emit ionising photons \citep{Geil17}. 21cm images around these objects would show an ionised region approximately centered on the bright galaxy. If many of these images were stacked, we would see an \textit{average} ionised region around the brightest galaxies. These stacks will provide an alternative detection strategy which could increase the sensitivity of observations, allowing us to probe further into the EoR and understand the nature of the galaxies that reionised the Universe.

In this work, we investigate stacking of 21cm intensity maps around bright galaxies using the reionization history and galaxy population from the \textsc{Bluetides} cosmological simulation. We draw connections between the appearance of these stacked profiles, the properties of the sources they are centred on, and the distribution of ionised hydrogen in the early Universe. Using these relations from the simulation, we predict the signal for upcoming galaxy samples from the \textit{Wide-Field Infrared Survey Telescope} (WFIRST) \textit{High Latitude Survey} (HLS) and observations of the redshifted cosmological 21cm signal from the low-frequency Square Kilometer Array (SKA1-low).

This strategy was proposed in \citet{Geil17}, where analytic spherical HII regions were constructed from simulated luminosity functions and simulated relationships between HII region size and UV luminosity. We expand on this idea in this work by directly stacking 21cm intensity maps in \textsc{BlueTides}, thereby including the effects of asymmetric HII regions, galaxies laying off-center relative to the HII regions they reside in, as well as the effect of radio foregrounds. 

This paper is laid out as follows: Section \ref{sec:method} covers the details of the \textsc{BlueTides} simulation, as well as the methodology for calculating the 21cm brightness temperature profiles and stacking them. Section \ref{sec:mocks} covers the details of our mock observations, including number of available galaxies and SKA1-low thermal noise. Section \ref{sec:results} shows our results, including how average HII regions are detected, and correlations with the bubble size distribution, galaxy luminosity, and spin temperature. Section \ref{sec:foregrounds} explores the effects of observational limitations on this method by including the effect of radio foregrounds. We conclude in Section \ref{sec:conclusion}.

\section{Simulation}\label{sec:method}
The \textsc{Bluetides} simulation \citep{YFeng16} is a cosmological hydrodynamic simulation within a cube of side length $400 h^{-1}$ comoving Mpc. It contains $2 \times 7040^3$ dark matter and baryonic particles of mass $1.2 \times 10^7 h^{-1} M_{\odot}$ and $2.36 \times 10^6 h^{-1} M_{\odot}$, respectively. Particle snapshots are available from $z=99$ down to $z=7$ \citep{Marshall19,Ni19}.

\textsc{Bluetides} uses a pressure-entropy formulation of Smoothed Particle Hydrodynamics \citep{Read10,Hopkins13}. This includes a multi-phase star formation model \citep{Springel03,Vogelsberger13}, cooling via radiative processes \citep{Katz96}, metal cooling \citep{Vogelsberger14}, the effects of molecular hydrogen on star formation \citep{Krumholz11}, a type II supernovae feedback model \citep{Okamoto10}, and a super-massive black hole model \citep{DiMatteo05}. \textsc{Bluetides} is the largest high-redshift hydrodynamic simulation run to date. The large volume allows us to have statistically significant samples of observable galaxies at the beginning of reionization. The WMAP9 cosmology \citep{Hinshaw2013} is used in \textsc{bluetides} and throughout this work: $\lbrace \Omega_{\rm m},\Omega_{\rm b},\Omega_\Lambda,h,\sigma_8,n_{\rm s} \rbrace = \lbrace 0.2814,0.0464,0.7186,0.697,0.820,0.971 \rbrace$.

\begin{figure}
\includegraphics[width=\linewidth]{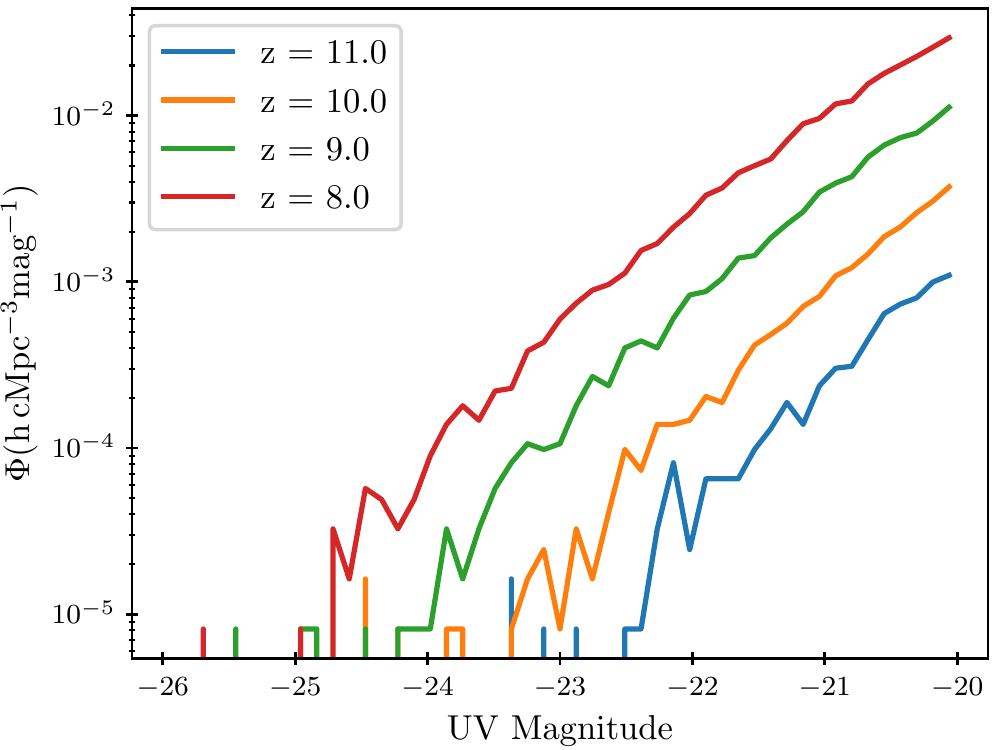}
\caption{UV Magnitude function predictions from \textsc{Bluetides}. Luminosity functions are shown at redshifts $z = $11, 10, 9 and 8.}
\label{fig:galhists}
\end{figure}

\begin{figure*}
\includegraphics[width=\linewidth]{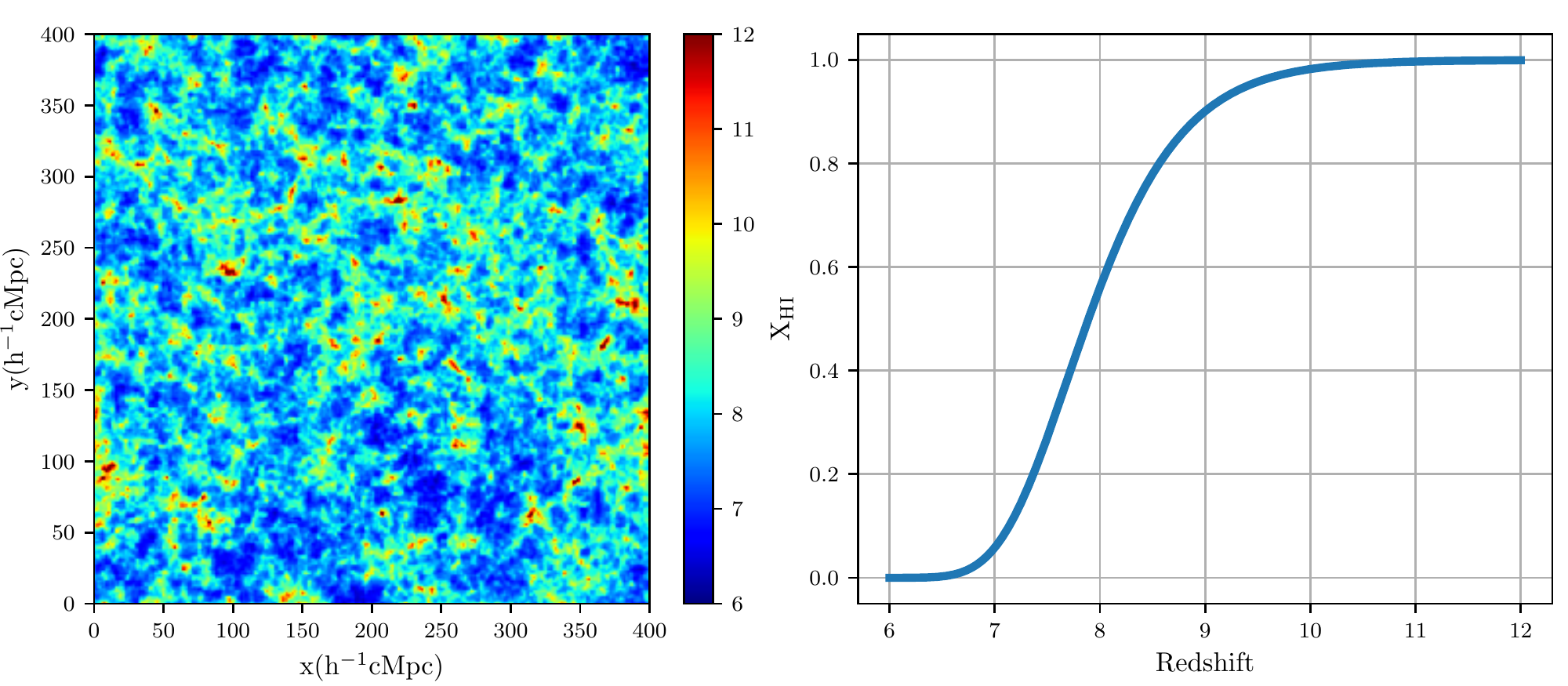}
\caption{Redshift of reionization slice (left) and global neutral fraction (right) in the \textsc{Bluetides} reionization scenario we use. The EoR has a median redshift of $z=8$, and a duration in redshift of roughly $\Delta z = 2$.}
\label{fig:xh}
\end{figure*}

In this work, we investigate how the properties of the brightest galaxies correlate with the structure of the 21cm signal. The UV magnitude distributions for these galaxies in \textsc{BlueTides} at redshifts 11, 10, 9, and 8 are shown in Figure \ref{fig:galhists}. These have been shown by \citet{YFeng16}, \citet{Waters16}, and \citet{Wilkins17} to be consistent with current observations. These distributions are used to estimate the number of galaxies per SKA field which are available for stacking.
The neutral fraction $x_{\rm{HI}}$ is set by the parametric reionization model used in \textsc{BlueTides} \citep{Battaglia2013}, which takes as input the density field at the chosen median redshift of reionisation $\bar{z}=8$  \citep[consistent with constraints from][]{Greig17}. A bias function is applied directly to the density field, which accounts for the correlation between the redshift of reionization and density fields on different scales\footnote{The models in \citet{Battaglia2013} were calibrated to reproduce reionisation in a set of $100 h^{-1} Mpc$ hydrodynamic and radiative transfer simulations with a particle mass of $2.58 \times 10^6 h^{-1} M_{\odot}$ and identified halos above $\sim 10^8 h^{-1}M_{\odot}$ on scales larger than $1 \rm{h^{-1} Mpc}$.} mitigating the need to identify halos during the EoR, the smallest of which which are not resolved in \textsc{Bluetides}. When applied to the \textsc{BlueTides} density field, this model results in reionization redshift ($z_{\rm{reion}}$) field in a $400^3$ ($1 \, \rm{h^{-1} cMpc}$ resolution) grid. The neutral fraction in each cell ($x_{\rm{HI}}$) at a particular redshift $z$ is set to unity in cells where $z_{\rm{reion}} < z$, and zero otherwise. The global neutral fraction and topology of reionization in \textsc{BlueTides} are shown in Figure \ref{fig:xh}.

As noted above, this method does not directly take into account source properties, which may contribute to the positions of the brightest galaxies within HII regions, and the asymmetry of those regions. It is possible that a model that computes ionisation structure directly from emitted ionising photons would result in early HII regions being more spherical, and centred closer to the brightest galaxies within them \citep[e.g:][]{Geil17}, making them easier to detect in stacked images.

\begin{figure*}
\includegraphics[width=\linewidth]{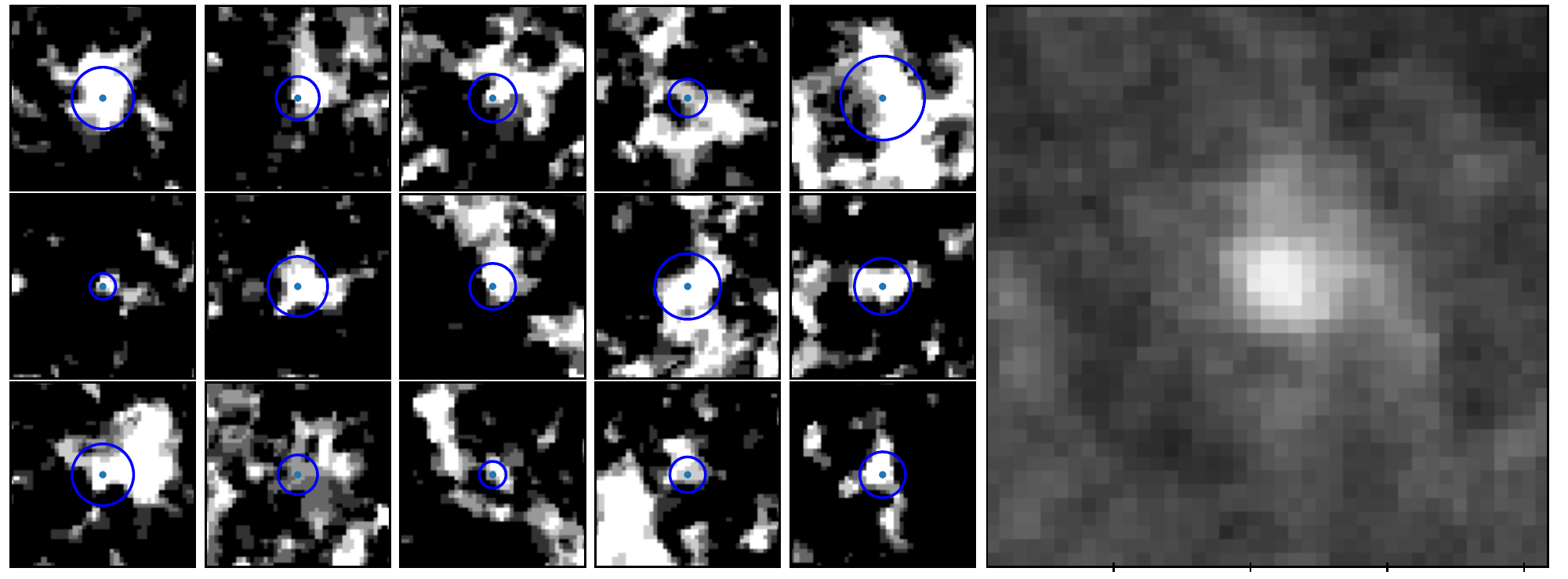}
\caption{The brightest 15 galaxies at redshift 9, $x_{HI} = 90.1\%$, the surrounding ionisation fields, and estimated individual bubble sizes from ray-tracing (shown as circles centred on each galaxy). The dimensions of each slice are 40 x 40 x 5 $\rm{h^{-1}Mpc}$. The color represents the average volume-weighted neutral fraction along the 5 $\rm{h^{-1}Mpc}$ depth of each slice. We also show the stacked profile of these 15 ionised regions in the rightmost panel, illustrating how stacking can smooth over the asymmetry of individual HII regions}
\label{fig:tiles}
\end{figure*}

One caveat to this model is that partially ionised structures on scales smaller than 1  $\rm{cMpc h^{-1}}$ are not considered, due to the fact that cells are either entirely ionised or entirely neutral. As a result, ionised regions could have less well-defined boundaries in reality than in our analysis, meaning the stacked images will be smoothed out, and possibly harder to detect. This will have the greatest effect when smaller HII regions (a few Mpc across) are stacked. However, the effect of the binary ionisation state will be somewhat minimized due to the fact that we will be stacking 21cm images around the brightest galaxies at a particular redshift, which we expect to be inside HII regions at least a few Mpc across (see Figures \ref{fig:tiles}, \ref{fig:correlations}). Ionised regions of radius $\sim 1 \rm{cMpc h^{-1}}$ are unlikely to be detected with our stacking pipeline due to redshift uncertainty, even if partially ionised cells were correctly represented in our model.

Further caveats to this model include the fact that the reionization grid is applied in post-processing, meaning inhomogeneous feedback from reionization is not included. However, since we are studying the earliest phases of reionization, around the brightest galaxies in the simulation which are well above the Jeans mass, we do not expect reionization feedback to have a large effect.

\begin{figure}
\includegraphics[width=\linewidth]{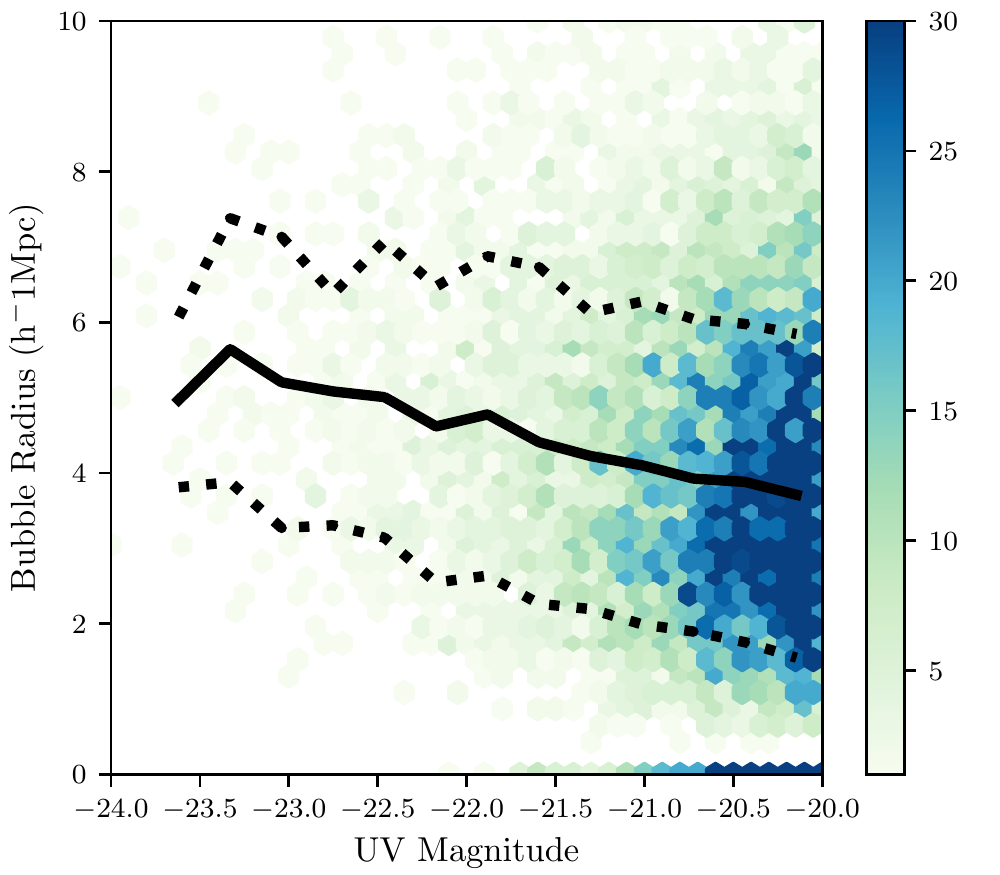}
\caption{HII region radius distributions, binned by the UV magnitude of the brightest galaxy within them, at redshift 9. The solid black lines show mean bubble radius in each bin, and the dotted lines show the 16th and 84th percentiles, such that the central region contains 67\% of HII regions in each bin. We see that dimmer galaxies generally host smaller ionised regions. There is also an increasing population of galaxies outside a HII region (with zero bubble radius) as UV magnitude increases.}
\label{fig:Rvsmag}
\end{figure}

Estimated bubble sizes for the brightest 15 galaxies at redshift 9, where the global volume-weighted neutral fraction is 90.1\%, are shown in Figure \ref{fig:tiles}. Although these are the brightest galaxies in the simulation, they do not necessarily reside in the centre of the largest HII regions in the simulation. Many of these bright galaxies are off-centre in these regions, and many have larger HII regions nearby that the galaxy itself is not a part of. We calculate an estimate of the true size of HII regions in the simulation via a ray-tracing method, taking the average distance from each galaxy to the nearest ionisation boundary (where $X_{HI} < 0.5$) in $1000$ randomly selected directions \citep[e.g:][]{Geil17}. The distribution of HII region sizes, calculated via this ray-tracing method, plotted against the UV magnitude of the brightest galaxy within them is shown in Figure \ref{fig:Rvsmag}.

We calculate the 21cm brightness temperature in the \textsc{Bluetides} volume using \citep{Furlanetto06}:
\begin{multline}\label{eq:dtb}
    \delta T_{b,21} = 27 K \left( \frac{\Omega_b h^2}{0.023} \right)
    \left( \frac{0.15}{\Omega_m h^2}\frac{1+z}{10} \right) ^{1/2} \\
    \left( 1-\frac{T_{CMB}}{T_s} \right) (1+\delta)x_{HI}
\end{multline}
where $\Omega_{b}$ and $\Omega_m$ are the abundances of baryons and matter respectively, $T_s$ is the 21cm spin temperature, $T_{CMB} = 2.73(1+z)$ is the temperature of the cosmic microwave background, and $x_{HI}$ is the local neutral fraction.

Unless otherwise stated, we assume the IGM is `saturated', where the spin temperature term in the above equation $\frac{T_{CMB}}{T_s} \approx 0$, and the IGM is always seen in net emission of 21cm radiation. In section \ref{sec:EOS}, we examine an `unsaturated' case which follows the global spin temperature evolution of the `bright galaxies' scenario from the Evolution of 21cm Structure (EOS) simulations \citep{EOS2016}, which has a similar neutral fraction evolution to our reionization model. Using this spin temperature evolution, at redshifts 11,10, 9 and 8 the brightness temperature of neutral hydrogen at mean density is -98.6, -41.6, -1.06 and 18.7 mK respectively.

\subsection{Stacking}\label{sec:stackmethod}
We centre mock $\delta T_b$ observations around galaxies in \textsc{BlueTides}. The centre point of each image is slightly displaced from the galaxy along the line-of-sight to model redshift uncertainties (see section \ref{sec:ngal}).
Each grid cell is then assigned a redshift depending on its distance from this point along the line-of-sight direction, where the central point is always at the redshift of the simulation snapshot. 21cm grids are then stacked in $80 \times 80 \times 80$ $\rm{h^{-1} Mpc}$ sub-cubes, centred on the closest cell to the central point, so that grid cells are aligned.

While the reionization model produces asymmetric HII regions, even towards the beginning of reionization, we expect a signal that is roughly cylindrically symmetric after stacking, since the amount and direction of asymmetry will be uncorrelated between each HII region. The right panel of figure \ref{fig:tiles} shows the HII profile of a region stacked around the brightest 15 galaxies in \textsc{BlueTides}. Along with the asymmetry of the ionised regions, the mis-centreing of bright galaxies within HII regions is the largest source of error when inferring statistics of the bubble size distribution from stacked profiles. Both of these effects will smooth out the stacked profiles depending on the average amount of asymmetry or the average distance of each galaxy from the centre of its HII region. However, the stacked profile itself will still be approximately symmetrical.

\section{Mock observations}\label{sec:mocks}
\subsection{Mock Surveys}\label{sec:ngal}
The signal for stacked 21cm images will be sensitive to the number of visible galaxies in the SKA field. The number of galaxies that we would expect to find in a survey can be calculated from the \textsc{bluetides} luminosity function and the SKA survey volume, determined by central redshift $z$, depth $\Delta z$ (such that the redshift boundaries are at $z \pm \Delta z / 2$), and the field of view $\Omega _{SKA} = \frac{\pi}{4} \left( \frac{\lambda}{D} \right) ^2$. Here $\lambda$ is the observed wavelength of rest-frame 21cm radiation at the redshift of observation, D is the station diameter of 35m. At redshifts $z = \lbrace 11,10,9,8 \rbrace$, the field of view is $\Omega_{SKA} \approx \lbrace 13.4, 11.2, 9.28, 7.52 \rbrace$ square degrees respectively.

\begin{figure}
\includegraphics[width=\linewidth]{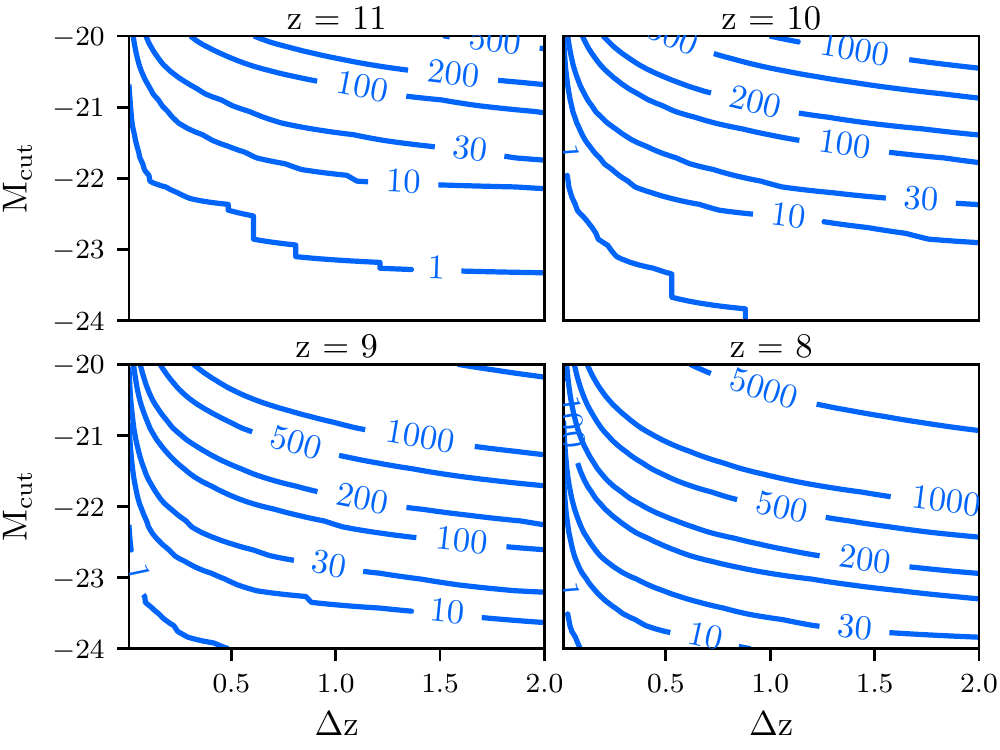}
\caption{Number of expected galaxies brighter than a UV magnitude $M_{\rm{cut}}$ within a volume of a survey with the SKA field of view and depth $\Delta z$. Contour lines are drawn at 1, 10, 30, 100, 200, 500, 1000, and 5000 galaxies.}
\label{fig:ngals}
\end{figure}

The predicted number of galaxies brighter than a limiting UV magnitude $M_{\rm{cut}}$ at redshift $z$ in the SKA field of view with depth $\Delta z$ is shown in Figure \ref{fig:ngals}. At redshift 9 and with a depth of $\Delta z = 1$, we would expect to have 1112 galaxies brighter than  $M_{cut} = -20.75$ in the volume. Magnitude cuts are chosen based on the WFIRST HLS $5\sigma$ limit at the redshift upper bound in the chosen volume, and those cuts at redshifts $z = \lbrace 11,10,9,8 \rbrace$ are $M_{cut} = \lbrace -21.03, -20.90, -20.75, -20.59 \rbrace$ respectively. At redshifts $z = \lbrace 11,10,9,8 \rbrace$ the number of galaxies above the magnitude cut in the \textsc{bluetides} volume that we sample from is 132, 633, 3241, and 12837, and the number of sampled galaxies within a SKA field of view is 54, 239, 1112, and 3936 respectively.

The instrument used to observe these galaxies is also important in providing accurate locations, particularly along the line-of-sight. Uncertainty in the positions of bright galaxies smooths out brightness temperature profiles when stacked, resulting in shallower signals that are harder to detect. WFIRST grism spectroscopy will have an associated redshift error of $\sigma_z = 0.001 (1+z)$ \citep{Spergel15}. At redshift 9, this results in a line-of-sight position uncertainty of $\sigma_d = 2.56 \, \rm{h^{-1} cMpc}$. Angular position uncertainty is assumed to be negligible.

\subsection{Instrument Sensitivity}\label{sec:noise}
In order to create mock observations, we use the package \textsc{21cmSense}\footnote{https://github.com/jpober/21cmSense} \citep{Pober2013,Pober2014} to create realisations of thermal noise for the SKA1-low. \textsc{21cmsense} takes the antenna configuration of SKA1-low\footnote{The antenna layout used is the \href{https://astronomers.skatelescope.org/wp-content/uploads/2015/11/SKA1-Low-Configuration_V4a.pdf}{SKA1-low Configuration V4a}}, along with the system temperature and dish size, to calculate sensitivity to line-of-sight and perpendicular modes at a certain frequency. We use the 21cm rest-frame frequency at the snapshot redshift as our central frequency, and generate 2D sensitivity power spectra using \textsc{21cmsense}. Assuming cylindrical symmetry, we interpolate the 2D power spectra to the same modes sampled by our brightness temperature grids, generating fluctuation amplitudes for each point in Fourier space for a 1080 hour observation (more detailed parameters used in the package can be found in Table \ref{tab:cmsense}). This is then added, with a random phase for each mode (assuming uncorrelated noise between modes), to the Fourier transform of the brightness temperature grids from \textsc{BlueTides} to simulate thermal noise from SKA1-low.

\begin{table}
\begin{center}
\begin{tabular}{ |c|c|c|c| }
 Parameter & Value \\
 \hline
 \hline
 Integration time & $60 \rm{s}$ \\
 \hline
 Observing days & 180 \\
 \hline
 Hours per day & 6 \\
 \hline
 Station diameter & 35m \\
 \hline
 Reciever temperature & 50K \\
 \hline
 Number of Antennas & 513 \\
 \hline
 Longest baseline used & 10km \\

\end{tabular}
\end{center}
 \caption{Observing parameters to calculate the thermal noise for SKA1-low with \textsc{21cmSense}} \label{tab:cmsense}
\end{table}

\begin{figure}
\includegraphics[width=\linewidth]{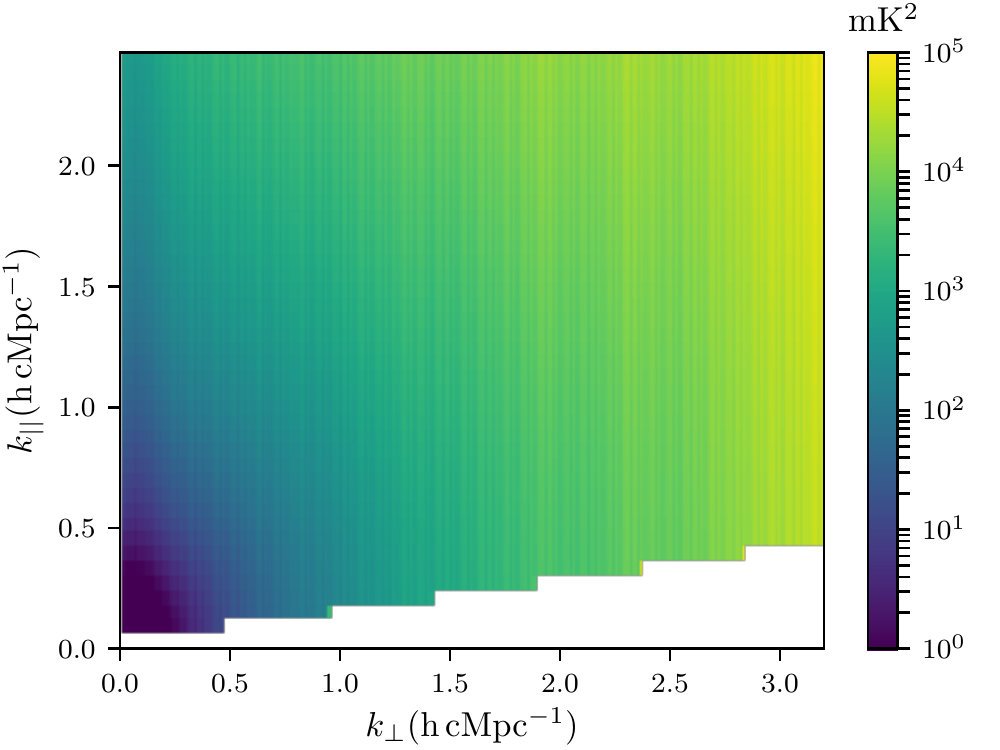}
\caption{21cm thermal noise of SKA1-low on sampled scales above $1 \rm{cMpc^{-1}h}$. Produced by the package \textsc{21cmcense} at redshift 9.}
\label{fig:cmsense}
\end{figure}

Regions of Fourier space in the simulation that are not probed by the instrument are discarded in both the image and noise cubes, including the $k = 0$ modes. We make additional cuts in Fourier space above $k_{\perp} = 1.5 \rm{h \, cMpc^{-1}}$ and $k_{\parallel} = 1.5 \rm{h \, cMpc^{-1}}$. This acts as a low-pass filter which removes the higher small scale noise while retaining most of our signal, which will usually be several $\rm{h \, cMpc^{-1}}$ wide due to the size of the HII regions. This filter smooths over both perpendicular and line-of-sight modes, since we expect the signal to be roughly symmetrical, however in practice taking different $k_{\parallel}$ cuts makes little difference to our results. Results in Section \ref{sec:results} assume perfect subtraction of radio foregrounds. In Section \ref{sec:foregrounds} we show results including model foregrounds by excluding areas of Fourier space below a certain line \citep[e.g:][]{Hassan19}.
Figure \ref{fig:cmsense} shows the 2D sensitivity power spectrum under the parameters laid out above, before cuts to Fourier space are made.

\section{Results}\label{sec:results}
\begin{figure}
\includegraphics[width=\linewidth]{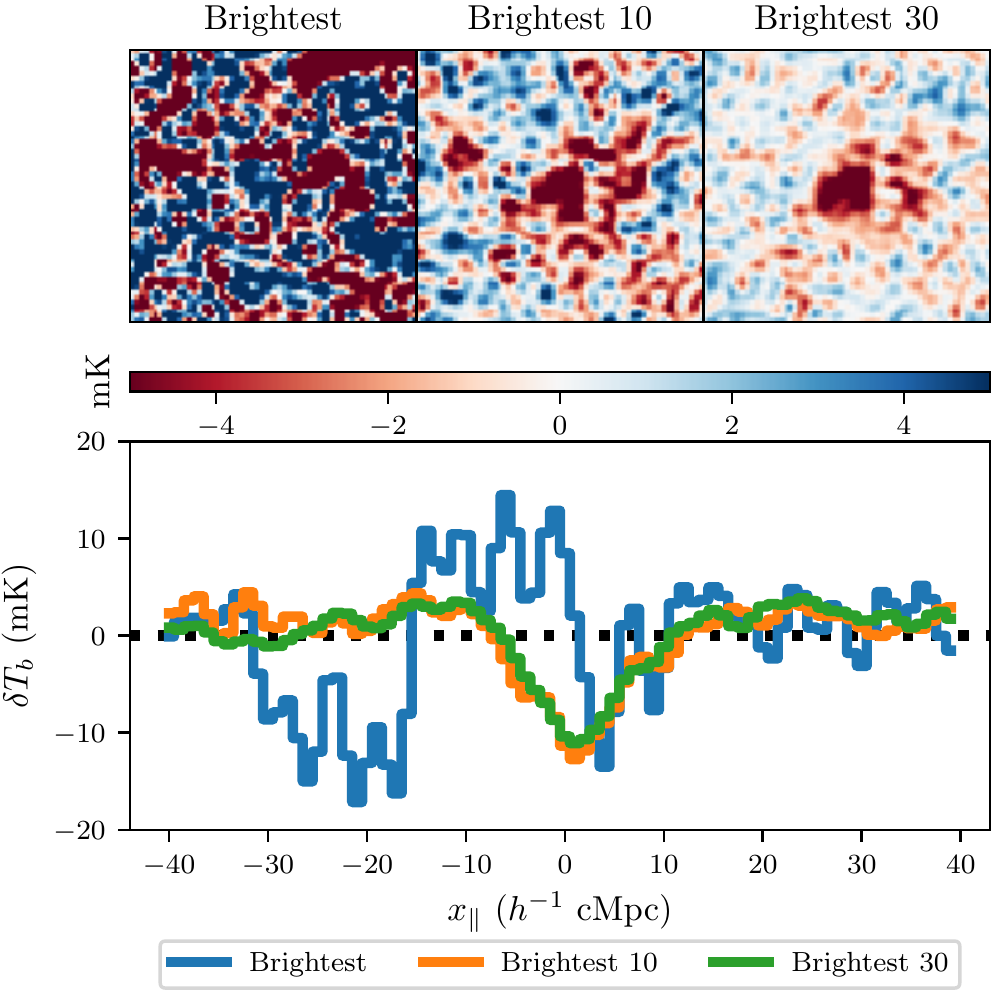}
\caption{Top: $60 \times 60 \times 5 \rm{cMpch}^{-1}$ 21cm brightness temperature slices of the single (left), 10 (middle), and 30 (right) brightest galaxies in our sample at redshift 9. The central region becomes clearer as more images are stacked.
bottom: A 1-Dimensional view of the same galaxies as the top panel, showing the brightness temperature through the line-of-sight of the galaxy averaged along a $5 \times 5 \, \rm{cMpch}^{-1}$ column.}
\label{fig:single_stack}
\end{figure}

We begin by considering the stacks of brightness temperature fields around up to 128 galaxies in the \textsc{Bluetides} simulation. We build a sample of galaxies by first estimating the number of galaxies brighter than the WFIRST HLS $5\sigma$ UV magnitude limit $(M_{\rm{cut}} = -20.75)$, in a volume determined by a central redshift $z=9$ with depth $\Delta z = 1$, and the field of view for the Square Kilometer Array (9.28 square degrees at $z = 9$). This results in 1112 galaxies. This number of galaxies are sampled from the \textsc{Bluetides} galaxies at the central redshift and above the same magnitude cut with replacement, representing a sample of available galaxies in an SKA1-low field. The regions around up to the brightest 128 of these galaxies are then selected as our sample for stacking. We take a random rotation and reflection along two axes, and keep the line-of-sight axis constant for the purposes of adding redshift uncertainty (this gives us 8 possible configurations per galaxy in \textsc{Bluetides}). The top panel of Figure \ref{fig:single_stack} shows a comparison of the brightness temperature around the brightest galaxy in one sample compared to a stack of the 10 and 30 brightest galaxies in the same sample. Given noise, redshift errors, asymmetry and mis-centreing of the ionised region as shown in figure \ref{fig:tiles}, distinguishing a single ionised region from the background could be difficult for some galaxy samples. However stacking many regions diminishes background fluctuations, asymmetries and noise, resulting in an average bubble profile which is roughly Gaussian.
As we increase the number of galaxies stacked, background variation decreases and the average profile converges. The bottom panel of Figure \ref{fig:single_stack} shows the decrease in background noise and fluctuations as we stack greater numbers of images along the line of sight spectra averaged within a 5 $\rm{h^{-1}cMpc}$ wide aperture.

\subsection{Detecting HII regions in stacked images}\label{sec:single}
The average ionised region profile is fit to a 3D Gaussian with separate widths along the line-of-sight and sky plane, using a Monte-Carlo Markov Chain fitting method of the form,
\begin{equation}\label{eq:gauss}
    f(\rho,z) = A e^{-\left( \frac{\rho^2}{2 \sigma_\rho^2} + \frac{z^2}{2 \sigma_z ^2} \right)},
\end{equation}
where $\rho$ is the transverse distance from the centre of the image, and $z$ is the line-of-sight distance.

We choose $A$ as the parameter describing the depth (or height) of the profile, and $\sigma_\rho$ as the parameter describing its width. We require separate widths along the two dimensions in order to fit the profiles because redshift uncertainty will elongate the profile along the line-of-sight. We note that $\sigma_z$ is not truly a free parameter, as it results from the convolution of a spherically symmetric Gaussian with the redshift uncertainty. However, we take the conservative approach of allowing it to take any value in our fits. It also becomes necessary to fit separately to $\sigma_z$ when modelling foreground contamination, where the small-scale sky-plane structure and large-scale line-of sight structure is removed, changing the shape of the profiles. Since $\sigma_z$ contains no new information about the underlying HII structure however, it is not used for inference in this work. We do not fit to a background brightness temperature, since in our treatment of instrument noise, we remove the zero mode in Fourier space, so that the image effectively has zero mean.

We can look at the effect of stacking more quantitatively, in terms of the fit parameters $(A,\sigma_{\rho})$ and bootstrap sampling of the image stacks. We take 128 bootstrap samples of the same size, giving us different realisations of available galaxies, HII regions, and instrumental noise.
\begin{figure}
\includegraphics[width=\linewidth]{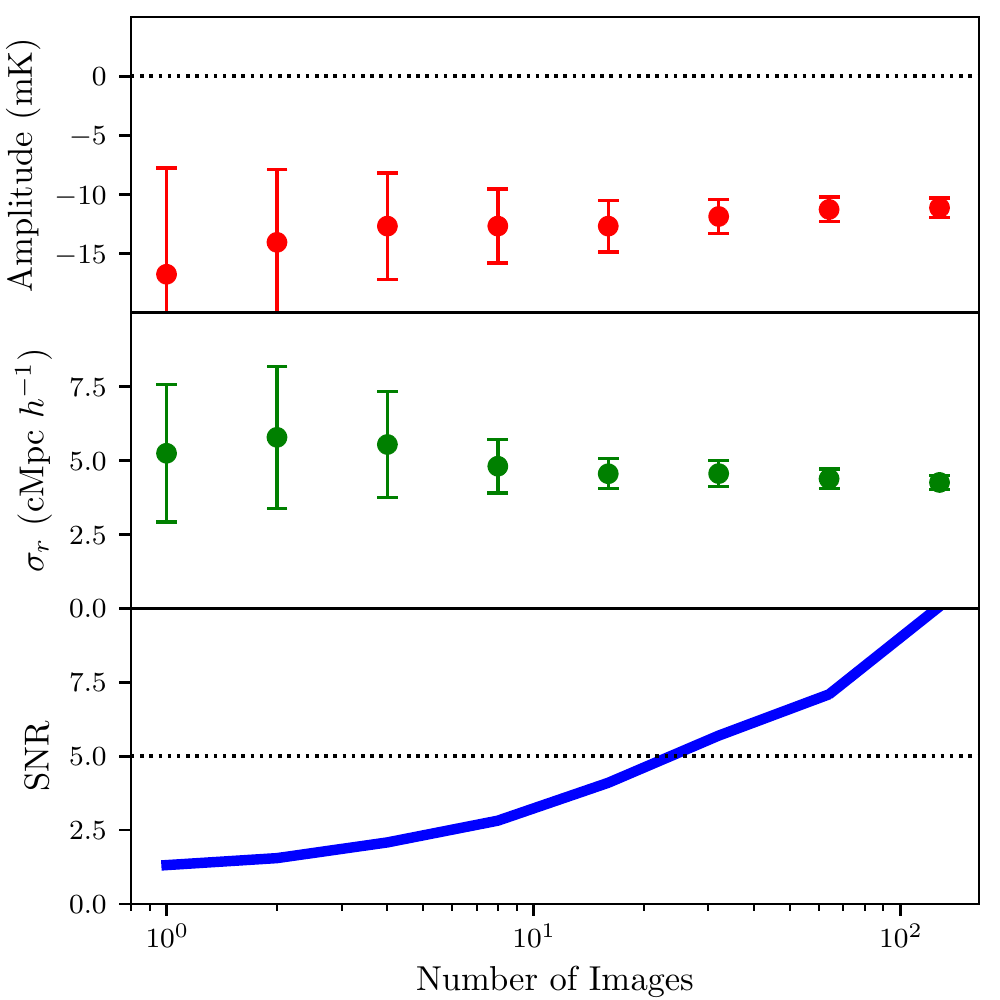}
\caption{Convergence of the Gaussian amplitude (top) and width (middle) versus number of stacked images, the amplitude becomes inconsistent with random pointings, where amplitude is fit closely to zero (see Appendix \ref{sec:rand}). The SNR (bottom) reaches 5 at $\sim30$ stacked images. Errorbars are 1$\sigma$.}
\label{fig:single_stack_pars}
\end{figure}
We choose the depth of the stacked absorption to measure the strength of the signal. Our Monte-Carlo fitting process gives us a mean $\mu_{A}$ and standard deviation $\sigma_{A}$ in depth. 
For comparion, we perform the same analysis on stacked images of random locations in the \textsc{BlueTides} volume. We find that amplitudes are distributed closely around zero, and widths are widely distributed around the scale of density fluctuations and instrument noise (see appendix \ref{sec:rand} for distributions of parameter for random pointings). Since this wide range includes the width of many HII regions, we only use the amplitude as a metric of detection.

We calculate the signal to noise ratio (SNR) by comparing the distributions of fitted depths between the stacked images of galaxies and random pointings. We find that both distributions are roughly Gaussian, so we characterise them by their means $\mu_{\rm{A}} , \mu_{\rm{R}}$ and standard deviations $\sigma_{\rm{A}} , \sigma_{\rm{R}}$ where the subscript A refers to the galaxy case, and R to the random pointings.

The signal to noise radio in this case is the mean of the distribution of differences in fit amplitude between galaxy and random pointings, divided by the standard deviation of this difference:
\begin{equation}\label{eq:snr}
    \rm{SNR} = \frac{|\mu_{diff}|}{\sigma_{diff}} \approx \frac{|\mu_R - \mu_A|}{\sqrt{\sigma_R^2 + \sigma_A^2}}.
\end{equation}
Figure \ref{fig:single_stack_pars} shows the convergence of our Gaussian fit parameters as more images are stacked. The stacked images reach a SNR greater than $5$ at $30$ images.

We find that the SNR increases more slowly as the number of stacked images increases. This is because the smaller HII regions around dimmer galaxies are being added to the stack, resulting in a slightly shallower and narrower profile. Eventually, the SNR gain by reducing background fluctuations will be overcome by the decrease in fit amplitude caused by smaller ionised bubbles around dimmer galaxies being added to the stack. However, with <100 galaxies, we do not reach the turning point in SNR.

\subsection{Bubble Size Distribution}\label{sec:distribution}
\begin{figure}
\includegraphics[width=\linewidth]{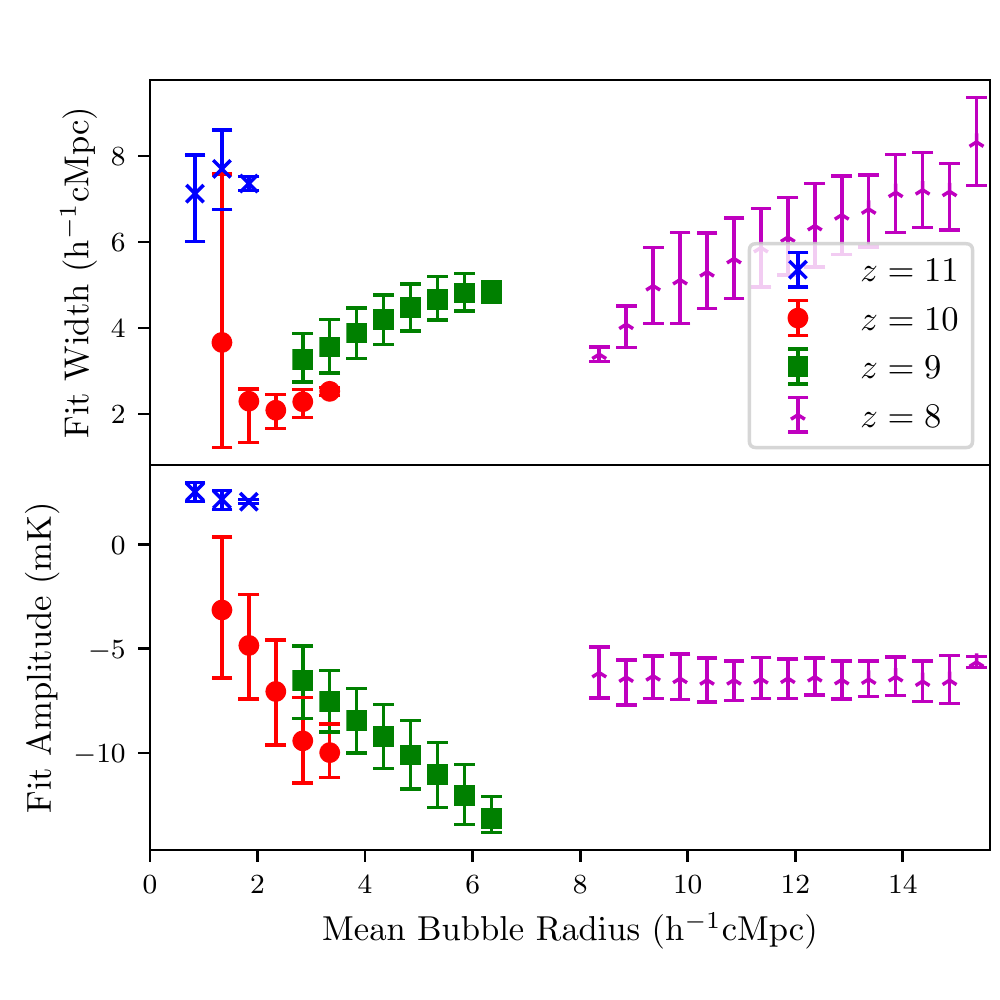}
\caption{Gaussian fit parameters versus mean HII region radius for various galaxy samples at $z = \lbrace 11,10,9,8 \rbrace$ within \textsc{BlueTides}. Top: Gaussian fit amplitude. Bottom: $1\sigma$ Gaussian width. There are clear relationships between the width and depth of the fits and the bubble sizes.}
\label{fig:correlations}
\end{figure}

We next investigate the relationship between our fit parameters and the properties of the HII regions within each stack of 30 galaxies using the same bootstrap sampling method as section \ref{sec:single}. We include samples from redshifts $z=11$ $(X_{HI} = 99.7 \% , M_{cut} = 21.03)$, $z=10$ $(X_{HI} = 98.2 \% , M_{cut} = -20.90)$ $z=9$ $(X_{HI} = 90.1 \% , M_{cut} = -20.75)$ and $z=8$ $(X_{HI} = 66 \% , M_{cut} = -20.59)$ to test different stages of the EoR. We also create stacks drawn from all galaxies above the UV magnitude cut in each sample rather than the brightest, in order to test a wider range of HII region sizes at each redshift. Each stack of 30 images has a single fit depth and width, as well as a true mean bubble radius calculated from the ray tracing method described in section \ref{sec:method}. The correlations between these quantities are plotted in Figure \ref{fig:correlations}.

21cm images comprising of larger HII regions are generally fit to deeper and wider profiles. Above a mean bubble radius of $\sim 2 h^{-1}\rm{cMpc}$ there is roughly a linear correlation between each fit parameter and mean bubble radius, however the slopes, offsets and scatter in the correlations differ within each redshift group. In particular the $z=8$ group shows no correlation between fit amplitude and bubble radius, due to decreased mean density and overlap resulting in wide, shallow profiles that are less dependent on the galaxy at the centre of the image. Pearson correlation coefficients between each parameter at redshifts $z=8$, $z=9$, and $z=10$ are shown in Table \ref{tab:corr}. The slopes and y-intercepts of linear fits to each group are also presented in Table \ref{tab:corrfits}.
\begin{table}
\begin{center}
\begin{tabular}{ |c|c|c|c| }
 & $z=8$ & $z=9$ & $z=10$ \\
 \hline
 \hline
 Width & 0.479 & 0.467 & 0.163 \\
 \hline
 Depth & 0.0149 & -0.546 & -0.522 \\
 \hline
 
\end{tabular}
\end{center}
 \caption{Pearson linear correlation coefficients between fit parameters and HII region size} \label{tab:corr}
\end{table}

\begin{table}
\begin{center}
\begin{tabular}{ |c|c|c|c| }
 & $z=8$ & $z=9$ & $z=10$ \\
 \hline
 \hline
 Width slope ($\rm{h^{-1}cMpc}/\rm{h^{-1}cMpc}$) & 0.502 & 0.586 & 0.412 \\
 \hline
 Width y-int ($\rm{h^{-1}cMpc}$) & 0.155 & 1.63 & 1.15 \\
 \hline
 Depth slope ($\rm{mK}/\rm{h^{-1}cMpc}$) & 0.0139 & -1.87 & -5.33 \\
 \hline
 Depth y-int ($\rm{mK}$) & -6.60 & -1.15 & -5.26 \\
 \hline
 
\end{tabular}
\end{center}
 \caption{Linear fits to relations between fit parameters and HII region size} \label{tab:corrfits}
\end{table}

Below a bubble radius of $\sim 2 h^{-1} \rm{cMpc}$ the profiles become much wider and shallower, meaning they will be much harder to detect. The stacks around the smallest HII regions show a positive amplitude, meaning that the regions are too small to detect, or many of the galaxies in the stack are not in ionised regions at all. In this case the profiles are dominated by the dense neutral regions surrounding the galaxies rather than ionised bubbles, resulting in a fit approximately $\sim7 h^{-1}\rm{cMpc}$ wide with a positive height of $\sim +3$ mK.

While each redshift group shows a correlation between fit parameters and estimated bubble radius, there are many details of the simulation and reionisation model that can alter them. For example the positions of galaxies within HII regions, spin temperatures, and structure growth all have effects on fit parameters that depend on bubble size. However, if enough stacked images are available, one may be able to estimate a comparative bubble size between stacks. In order to obtain a calibrated relationship between these quantities, and estimate the bubble size distribution from stacked images, one would need to compare fit parameters for a range of galaxies against a suite of reionisation models.

\subsection{Correlation of Stacked Image Properties with UV Magnitude}
\begin{figure}
\includegraphics[width=\linewidth]{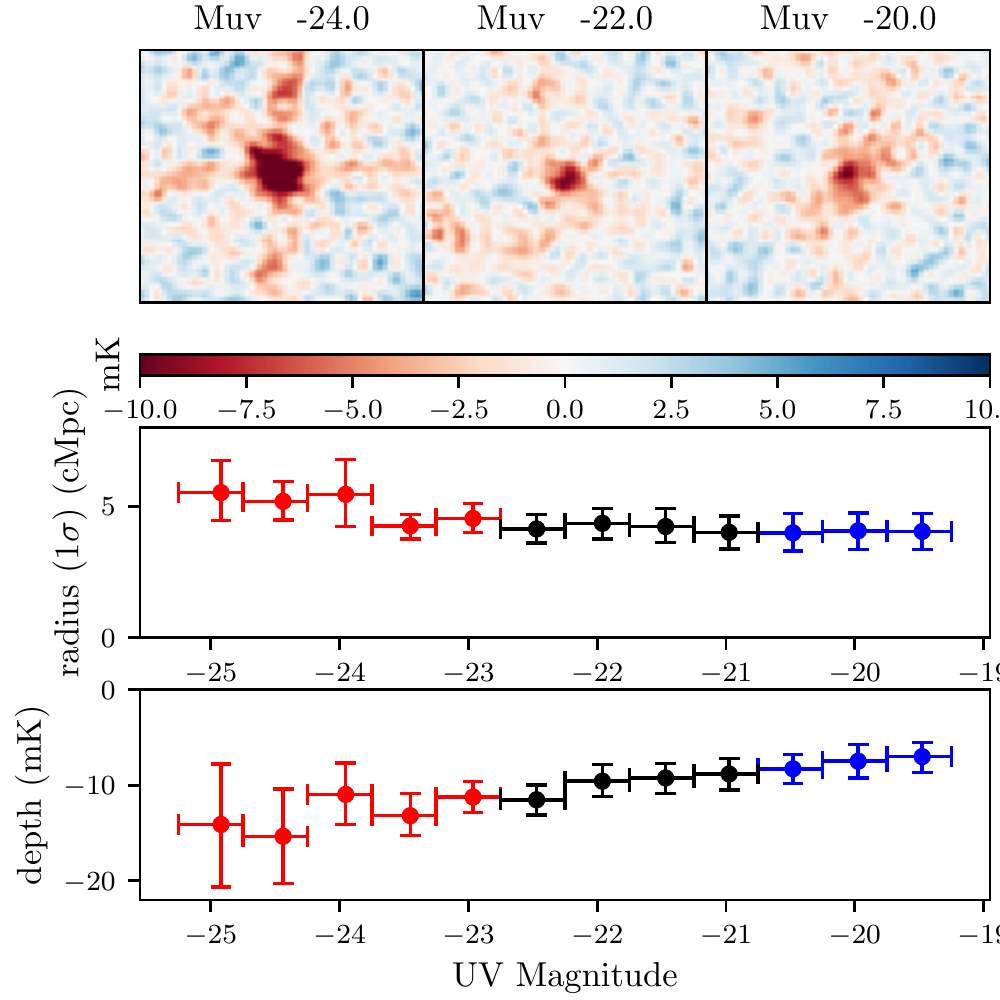}
\caption{Top: 21cm brightness temperature slices of 30 stacked galaxies, where the samples were chosen based on UV magnitude. Dimmer galaxies reside in visibly smaller HII regions.
Bottom: Gaussian fit parameters for binned galaxy samples. Brighter galaxies result in significantly deeper and slightly wider stacked profiles. Vertical errorbars are $1\sigma$ variation in the fit parameter, horizontal errorbars cover the magnitude bins, with points at the average UV magnitude in each bin. In both panels, magnitude bins below the WFIRST HLS $5\sigma$ limit are coloured blue, and bins where we do not expect more than 30 galaxies in the SKA1-low field of view are coloured red.}
\label{fig:mag_stack}
\end{figure}

This section details the correlations between galaxy magnitudes, and the resulting stacked 21cm brightness temperature profiles. We bin galaxies based on UV magnitude, and create stacked 21cm observations within each bin using the method discussed in section \ref{sec:stackmethod}. The resulting radii and depth of the profiles at redshift 9 are compared with galaxy UV magnitude in Figure \ref{fig:mag_stack}. In order to further explore the correlations between 21cm stacked images and UV magnitude of the galaxies in the simulation, we also display results for magnitude bins that are below the WFIRST HLS magnitude limit, as well as bins that we do not expect more than 30 galaxies in the SKA1-low field of view.

Our model shows a weak correlation between fit width and UV magnitude, and a moderate correlation between fit depth and UV magnitude. On average, galaxies of absolute UV magnitude -23 (-21) are hosted in ionised regions of radius 5.2 (4.1) Mpc which, when stacked with simulated SKA1-low thermal noise for a 1080h observation, are fit to Gaussian profiles of depth -11 (-9) mK and ($1\sigma$) width 4.5 (4.0) $\rm{h^{-1}cMpc}$ respectively, consistent with the results of figures \ref{fig:Rvsmag} and \ref{fig:correlations}. However, the scatter in the relationship between UV magnitude and fit width is very large compared to the range of widths across the UV magnitude bins. This is partially due to the large scatter in the $M_{\rm{UV}} - R$ relation in the simulation (see figure \ref{fig:Rvsmag}), as well as a decrease in SNR when stacking dimmer galaxies. The low-pass filter applied when adding thermal noise (see section \ref{sec:noise}) also has the effect of smoothing out smaller HII regions, however removing or extending this filter to smaller scales does not result in a stronger correlation, as the decrease in SNR overcomes the reduced amount of smoothing from the filter. The correlation of fit amplitudes with UV magnitude appears stronger, as the decrease in SNR, the smoothing from the low-pass filter, and the effect of smaller HII regions all result in a decrease in fit amplitude, and this should be taken into account when interpreting these correlations.

The correlations between galaxy properties and HII regions depend on the nature of reionisation. The contribution of different sized galaxies to reionisation and the correlation between the positions of galaxies and HII structure will change the strength of these relations. For example, a reionisation model dominated by brighter galaxies would show stronger correlation between UV magnitude and both fit parameters. This implies that stacked 21cm observations could be used to distinguish between different reionisation models. However, this will require sufficiently high SNR detections, so that the properties of the images correlate with the statistics of the EoR, and not with noise in the image.

\subsection{Spin Temperatures}\label{sec:EOS}
The spin temperature evolution is determined by the history of the first sources of Lyman alpha, UV and X-ray photons. In this section we show that stacked 21cm images could distinguish between positive and negative brightness temperature cases. We test two models of spin temperature. In the first model, the IGM spin temperature is saturated ($T_s >> T_{CMB}$ in equation \eqref{eq:dtb}), and the IGM is emitting at its maximum value. The second model follows the `Bright Galaxies' case presented in \citet{EOS2016}, where the IGM is seen in absorption for $z \geq 9$.

\begin{figure}
\includegraphics[width=\linewidth]{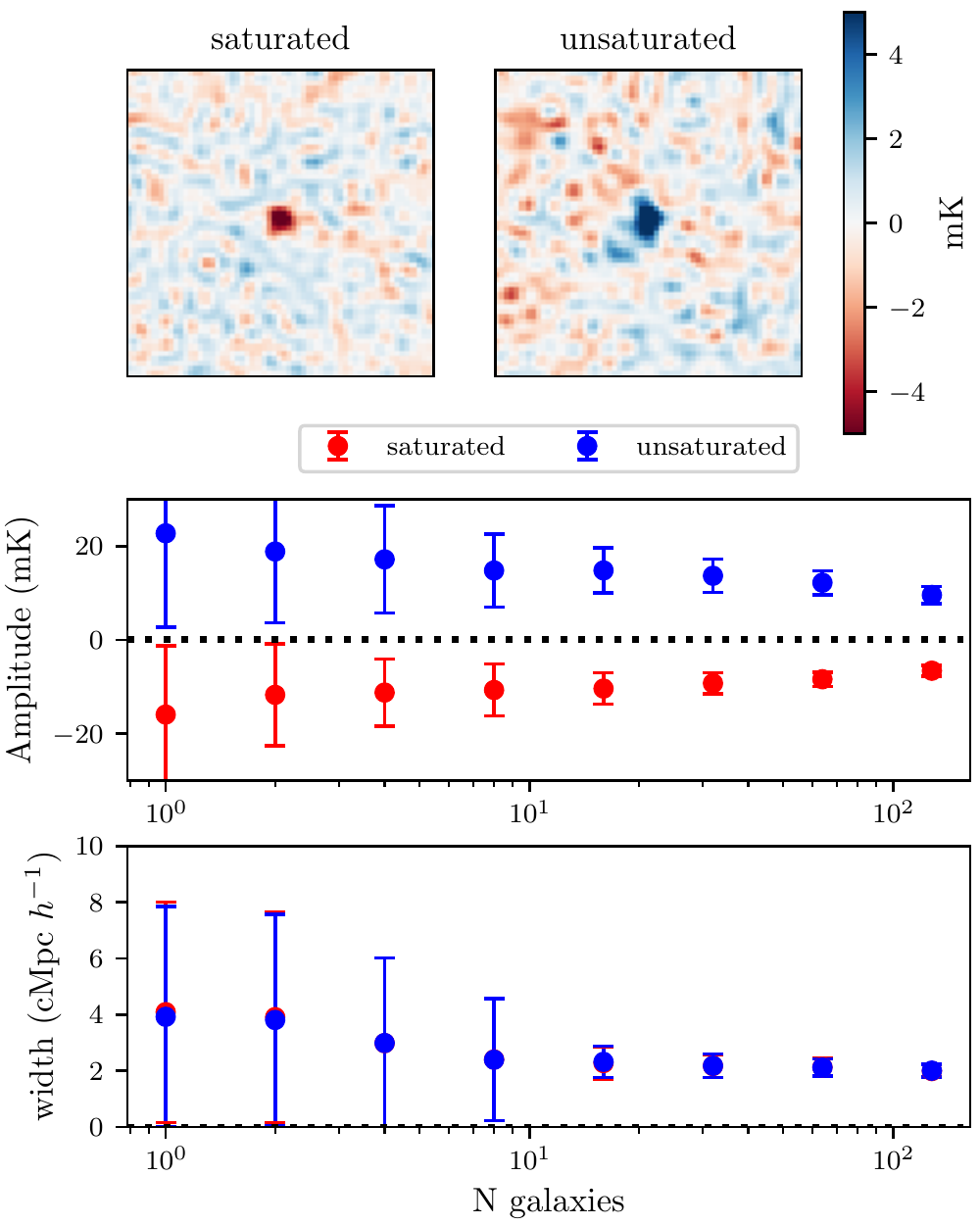}
\caption{Top: Stacked brightness temperature slices in a saturated (left) and unsaturated (right) IGM at $z=10$, where the central feature appears as a trough or peak in brightness temperature respectively.
Bottom: Gaussian fit parameters for stacks of up to 128 images, comparing the saturated and unsaturated cases. Distinguishing between these cases can be done with very few galaxies. The width of the stacked HII profiles are distributed almost identically in each case, however the profile depths have opposite sign.}
\label{fig:saturation}
\end{figure}

In Figure \ref{fig:saturation}, we show the slices and fit parameters in each spin temperature case, for up to 128 stacked images at redshift 10, using SKA1-low thermal noise described in Section \ref{sec:noise}. In the unsaturated case, the IGM is in net absorption of 21cm photons, and the brightness temperature of neutral hydrogen at mean density is -41.6 mK. Each stack is detected at over $5\sigma$ with $30$ stacked images, indicating that such observations could distinguish between these two cases.

The depth of the stacked profiles is linearly proportional to the spin temperature. As a result,  distinguishing between IGM in emission or absorption will be possible with these measurements. However, a more precise measurement of the spin temperature will be more difficult using this method, due to degeneracies with the distribution of sizes and shapes of early HII regions. For example, a decrease in fit amplitude may be due to smaller HII regions, more asymmetrical regions, an increase in instrumental noise or foreground levels, or a lower spin temperature.

\section{The Effects of Radio Foregrounds}\label{sec:foregrounds}
Radio interferometers measure the angular component of the 21cm signal in Fourier space, and it is therefore most natural to present observations of a 2D power spectrum, representing fluctuations on the sky, and along the line-of-sight, on separate axes. Since our stacked profiles are approximately Gaussian in real-space, so they are also approximately Gaussian in the 2D power spectrum, centred on zero. 

The EoR must be observed through radio foregrounds which outshine the EoR signal by several orders of magnitude \citep{Liu2014}. Fortunately foregrounds tend to be smooth in frequency, and occupy the low $k_{||}$ region of fourier space, although the inherent chromaticity of the telescopes push the foregrounds into higher $k_{||}$ regions, creating a foreground "wedge". One strategy is to avoid the foreground wedge when making 21cm observations \citep{Dillon14}. However removal of the foreground wedge will result in the loss of any signal inside of it.

\begin{figure}
\includegraphics[width=\linewidth]{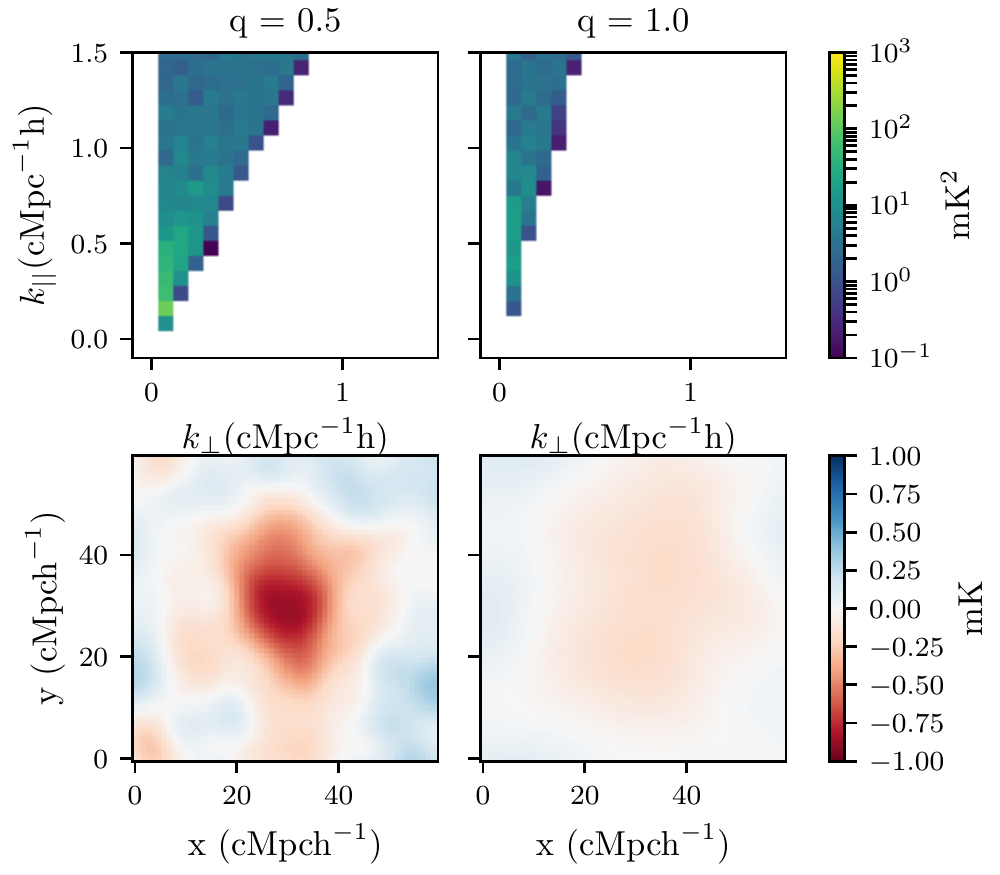}
\caption{Two foreground models at redshift 9, showing the signal in cylindrically averaged Fourier space (top), and slices through the centre of the stacked profile (bottom). We show results for maximum foreground contamination, $q=1$, and the case where foreground subtraction methods decrease the horizon slope by 50 \% $(q=0.5)$.}
\label{fig:fg_slice}
\end{figure}

To simulate foreground contamination in the wedge, we remove portions of Fourier space in our image below a line given by
\begin{equation}\label{eq:h1}
    k_{\parallel} \leq m k_{\perp}
\end{equation}
where
\begin{equation}\label{eq:horizon}
    m = \frac{DH_0E(z)sin(\theta)}{c(1+z)}.
\end{equation}
Here D is the comoving distance to the observation, $H_0$ is the Hubble constant, $E(z) \equiv \sqrt{\Omega_m(1+z) + \Omega_\Lambda}$, and $\theta$ is the beam angle. We take the horizon $\theta = \pi/2$ as a conservative limit. This is the maximum slope of the wedge for spectrally smooth foregrounds \citep{Liu2014}, which at redshifts $z = \lbrace 11, 10, 9, 8 \rbrace$ is $m = \lbrace 4.07,3.83,3.54,3.32 \rbrace$ respectively. Fourier mode bins that lie on this limit are down-weighted by a factor equal to the proportion of their volume above this line. We also consider models where the slope of the line is multiplied by a factor $0<q<1$, to represent scenarios where foreground subtraction or other techniques reduce the area of Fourier space dominated by foregrounds\footnote{A similar parameterisation of foreground contamination was used in \citet{Hassan19}} \citep[for examples of foreground mitigation techniques, see][]{Chapman19}. We show stacked profiles of 30 images at redshift 9, for $q = 1$ and $q = 0.5$ in Figure \ref{fig:fg_slice}. While a clear central trough in brightness temperature exists in the more optimistic case, it will be more difficult to detect ionised regions with maximal foreground contamination, where the signal to noise ratio is reduced to approximately 1.

\begin{figure}
\includegraphics[width=\linewidth]{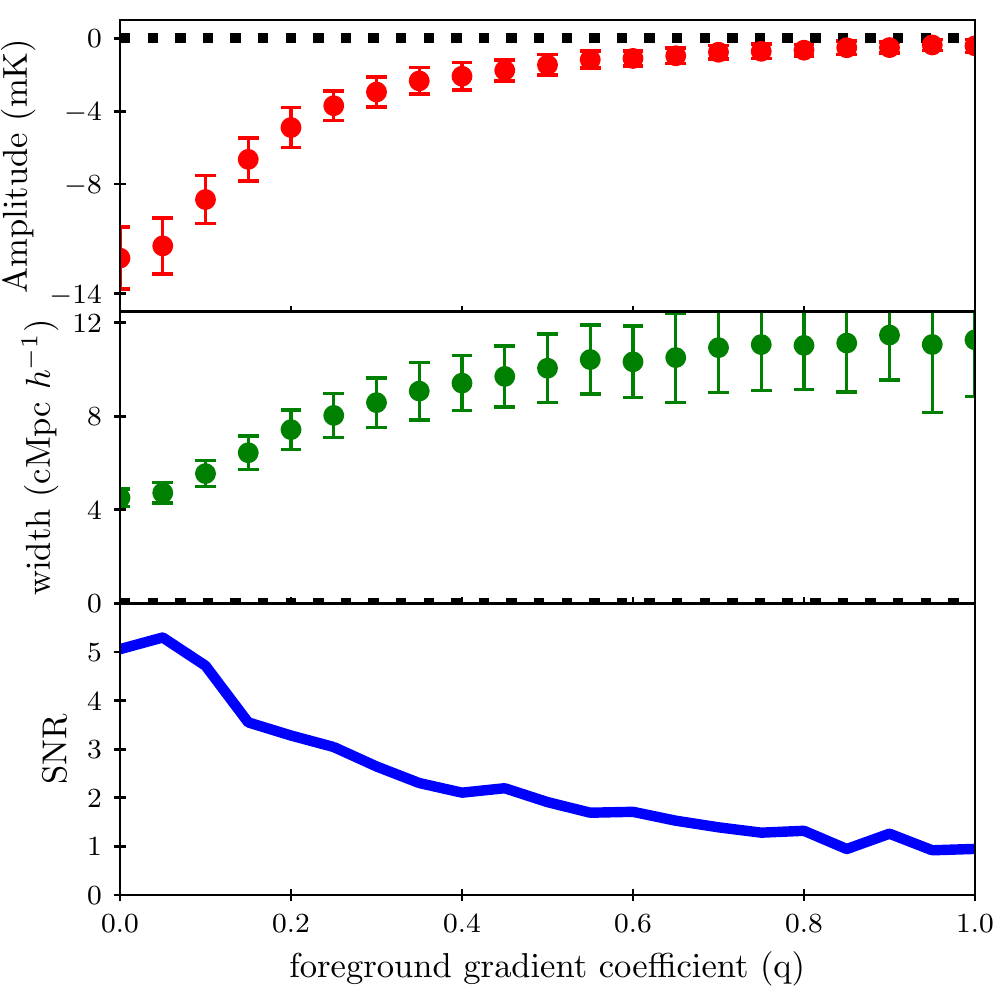}
\caption{The effect of radio foregrounds contaminating different regions of the 2D power spectrum. Each panel plots a Gaussian fit parameter versus the slope coefficient of the foreground wedge, where $q = 0$, represents the case of no contamination, and $q = 1$ represents the case of maximal foreground contamination for spectrally smooth foregrounds. Errorbars represent the standard deviation of the fit parameters. Top: Amplitude of the Gaussian fit. Center: $1 \sigma$ width of the Gaussian fit, Bottom: Average signal to noise for stacks at each foreground coefficient}
\label{fig:conv_fg}
\end{figure}

Since most of the signal power for an ionised region is at large scales in both perpendicular and parallel modes, the amount of signal rendered unusable by foreground contamination is roughly proportional to the slope of the wedge. Comparing Gaussian fits for foreground models with $0 \leq q \leq 1$ in Figure \ref{fig:conv_fg}, we can see that a significant amount of foreground mitigation will be required to detect early ionised regions with a stack of 30 redshifted 21cm images around bright galaxies.

\begin{figure*}
\includegraphics[width=\linewidth]{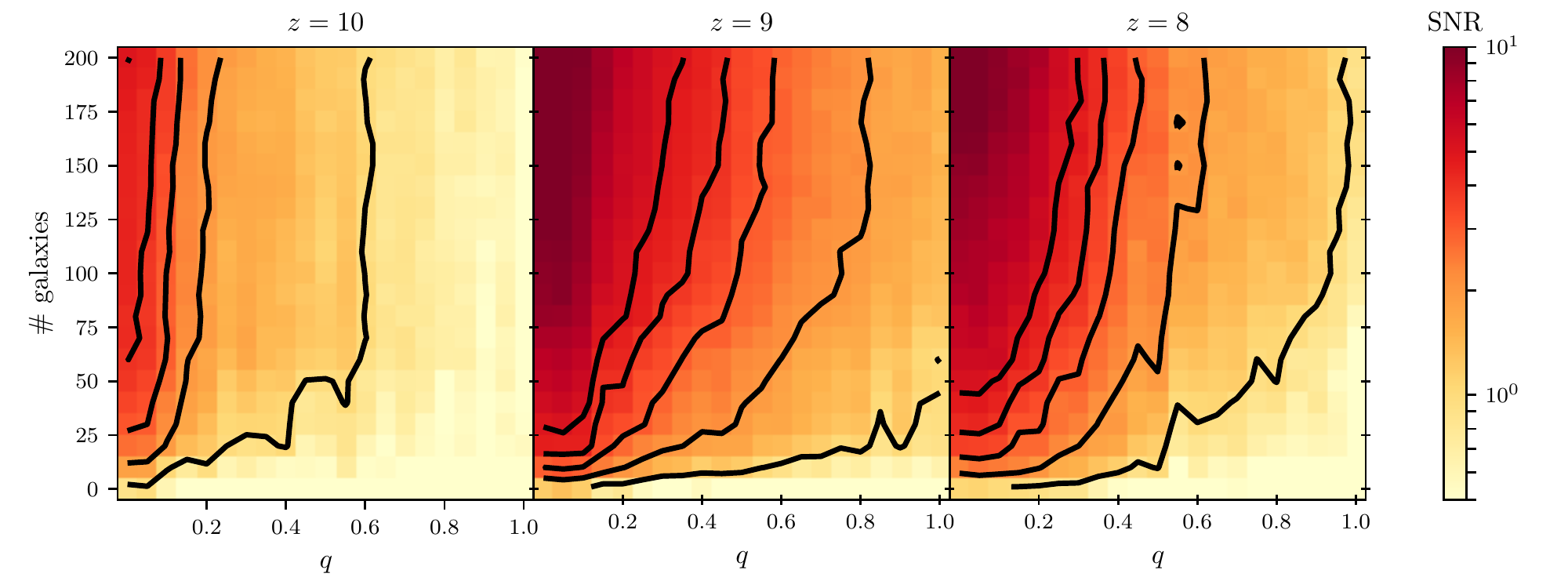}
\caption{Signal to noise ratio (SNR) versus foreground level and number of galaxies stacked at redshifts 10,9 and 8. Contours from right to left show $\rm{SNR} = (1,2,3,4,5)$}
\label{fig:snr}
\end{figure*}

Figure \ref{fig:snr} shows the signal to noise ratio (Equation \ref{eq:snr}) obtained when stacking larger numbers of galaxies, with differing levels of foreground contamination at redshifts 10, 9 and 8. Increasing the number of galaxies improves the SNR for low numbers of galaxies. However, as in the case of perfect foreground subtraction, beyond $\sim 100$ galaxies there are diminishing returns, because you will be stacking dimmer galaxies which add less to the signal.

Tentative detections $\rm{SNR} > 1$ could be achieved for early ionised regions at redshift $z=9$ with 100 galaxies and full foreground contamination. However, for a detection with SNR greater than 2, partial foreground subtraction will be required. An SNR of $(2,3,4,5)$ will require a fraction of foreground contaminated Fourier space of at most $q = (0.8,0.55,0.45,0.35)$ for less than 200 galaxies at redshift $z=9$.

\section{Conclusions}\label{sec:conclusion}
We have used the \textsc{Bluetides} hydrodynamic simulation to create mock observations of 21cm brightness temperature stacked around the brightest galaxies in the simulation. We have shown that stacking images of 21cm brightness temperature centred around bright galaxies provides a promising way of detecting the first large cosmological HII regions during reionization. Our main findings are as follows:

\begin{itemize}
    \item The detection of HII regions can be achieved with image stacks around relatively low numbers of bright galaxies. For example, in an idealised case with no foreground contamination, a $5\sigma$ detection could result from 30 galaxies at redshift 9.
    \item Stacked profiles are correlated with the HII region size distribution. With further modelling it could be possible to estimate a bubble size distribution from stacked profiles.
    \item By binning sources according to their brightness, an estimate of the relationship between UV magnitude and bubble size could be obtained. Our model shows little correlation between fit width and UV Magnitude, but shows that fit depth decreases with increasing UV magnitude. These correlations depend on the topology of reionisation, hence stacked profiles could be used to probe the contribution of galaxies of differing brightness to the EoR, provided enough images are obtained.
    \item The sign of the brightness temperature of the IGM affects the sign of the stacked profiles, which could be used to distinguish between IGM that is in net absorption or emission of 21cm photons.
    \item With enough stacked images, HII regions could be detected in the presence of foreground contamination. For example, if 100 21cm images are stacked around the brightest available galaxies in an SKA1-low field, a tentative $1.4\sigma$ detection could be made with full foreground contamination at redshift 9. However, if foreground subtraction reduces the contaminated region of Fourier space by 50 (80) percent, a 3 (5)$\sigma$ detection of early ionised regions can still be achieved with the same number of galaxies.
\end{itemize}

\section*{Acknowledgements}
JD is supported by the University of Melbourne under the Melbourne Research Scholarship (MRS).
This research was also supported by the Australian Research Council Centre of Excellence for All Sky Astrophysics in 3 Dimensions (ASTRO 3D), through project number CE170100013.
This work was performed on the OzSTAR national facility at Swinburne University of Technology. OzSTAR is funded by Swinburne University of Technology and the National Collaborative Research Infrastructure Strategy (NCRIS).
The BlueTides simulation was run on the BlueWaters facility at the National Center for Supercomputing Applications.
TDM and RACC are grateful for the hospitality of the Physics Department of the University of Melbourne, and for the support of  a Shimmins Fellowship (TDM) and Lyle Fellowships (TDM and RACC). 
TDM and RACC acknowledge support from  NASA NNX17AK56G, NASA ATP 80NSSC18K101 and NSF AST-1614853. RACC was also supported by NSF AST-1615940 and  NSF AST-1909193.

\section*{Data Availability}
The data underlying this article will be shared on reasonable request to the corresponding author. E-mail: daviesje@student.unimelb.edu.au



\bibliographystyle{mnras}
\bibliography{paper} 

\begin{thebibliography}{}
\makeatletter
\relax
\def\mn@urlcharsother{\let\do\@makeother \do\$\do\&\do\#\do\^\do\_\do\%\do\~}
\def\mn@doi{\begingroup\mn@urlcharsother \@ifnextchar [ {\mn@doi@}
  {\mn@doi@[]}}
\def\mn@doi@[#1]#2{\def\@tempa{#1}\ifx\@tempa\@empty \href
  {http://dx.doi.org/#2} {doi:#2}\else \href {http://dx.doi.org/#2} {#1}\fi
  \endgroup}
\def\mn@eprint#1#2{\mn@eprint@#1:#2::\@nil}
\def\mn@eprint@arXiv#1{\href {http://arxiv.org/abs/#1} {{\tt arXiv:#1}}}
\def\mn@eprint@dblp#1{\href {http://dblp.uni-trier.de/rec/bibtex/#1.xml}
  {dblp:#1}}
\def\mn@eprint@#1:#2:#3:#4\@nil{\def\@tempa {#1}\def\@tempb {#2}\def\@tempc
  {#3}\ifx \@tempc \@empty \let \@tempc \@tempb \let \@tempb \@tempa \fi \ifx
  \@tempb \@empty \def\@tempb {arXiv}\fi \@ifundefined
  {mn@eprint@\@tempb}{\@tempb:\@tempc}{\expandafter \expandafter \csname
  mn@eprint@\@tempb\endcsname \expandafter{\@tempc}}}

\bibitem[\protect\citeauthoryear{{Battaglia}, {Trac}, {Cen}  \&
  {Loeb}}{{Battaglia} et~al.}{2013}]{Battaglia2013}
{Battaglia} N.,  {Trac} H.,  {Cen} R.,   {Loeb} A.,  2013, \mn@doi [The
  Astrophysical Journal] {10.1088/0004-637X/776/2/81}, \href
  {https://ui.adsabs.harvard.edu/abs/2013ApJ...776...81B} {776, 81}

\bibitem[\protect\citeauthoryear{{Blyth} et~al.,}{{Blyth}
  et~al.}{2015}]{Blyth15}
{Blyth} S.,  et~al., 2015, in Advancing Astrophysics with the Square Kilometre
  Array (AASKA14). p.~128 (\mn@eprint {arXiv} {1501.01295})

\bibitem[\protect\citeauthoryear{{Chapman} \& {Jeli{\'c}}}{{Chapman} \&
  {Jeli{\'c}}}{2019}]{Chapman19}
{Chapman} E.,  {Jeli{\'c}} V.,  2019, arXiv e-prints, \href
  {https://ui.adsabs.harvard.edu/abs/2019arXiv190912369C} {p. arXiv:1909.12369}

\bibitem[\protect\citeauthoryear{{Datta}, {Bharadwaj}  \& {Choudhury}}{{Datta}
  et~al.}{2007}]{Datta07}
{Datta} K.~K.,  {Bharadwaj} S.,   {Choudhury} T.~R.,  2007, \mn@doi [\mnras]
  {10.1111/j.1365-2966.2007.12421.x}, \href
  {https://ui.adsabs.harvard.edu/abs/2007MNRAS.382..809D} {382, 809}

\bibitem[\protect\citeauthoryear{{Datta}, {Majumdar}, {Bharadwaj}  \&
  {Choudhury}}{{Datta} et~al.}{2008}]{Datta08}
{Datta} K.~K.,  {Majumdar} S.,  {Bharadwaj} S.,   {Choudhury} T.~R.,  2008,
  \mn@doi [\mnras] {10.1111/j.1365-2966.2008.14008.x}, \href
  {https://ui.adsabs.harvard.edu/abs/2008MNRAS.391.1900D} {391, 1900}

\bibitem[\protect\citeauthoryear{{Di Matteo}, {Springel}  \& {Hernquist}}{{Di
  Matteo} et~al.}{2005}]{DiMatteo05}
{Di Matteo} T.,  {Springel} V.,   {Hernquist} L.,  2005, \mn@doi [\nat]
  {10.1038/nature03335}, \href
  {https://ui.adsabs.harvard.edu/abs/2005Natur.433..604D} {433, 604}

\bibitem[\protect\citeauthoryear{{Dillon} et~al.,}{{Dillon}
  et~al.}{2014}]{Dillon14}
{Dillon} J.~S.,  et~al., 2014, \mn@doi [\prd] {10.1103/PhysRevD.89.023002},
  \href {https://ui.adsabs.harvard.edu/abs/2014PhRvD..89b3002D} {89, 023002}

\bibitem[\protect\citeauthoryear{{Feng}, {Di-Matteo}, {Croft}, {Bird},
  {Battaglia}  \& {Wilkins}}{{Feng} et~al.}{2016}]{YFeng16}
{Feng} Y.,  {Di-Matteo} T.,  {Croft} R.~A.,  {Bird} S.,  {Battaglia} N.,
  {Wilkins} S.,  2016, \mn@doi [\mnras] {10.1093/mnras/stv2484}, \href
  {https://ui.adsabs.harvard.edu/abs/2016MNRAS.455.2778F} {455, 2778}

\bibitem[\protect\citeauthoryear{{Furlanetto}, {Oh}  \& {Briggs}}{{Furlanetto}
  et~al.}{2006}]{Furlanetto06}
{Furlanetto} S.~R.,  {Oh} S.~P.,   {Briggs} F.~H.,  2006, \mn@doi [\physrep]
  {10.1016/j.physrep.2006.08.002}, \href
  {https://ui.adsabs.harvard.edu/abs/2006PhR...433..181F} {433, 181}

\bibitem[\protect\citeauthoryear{{Geil} \& {Wyithe}}{{Geil} \&
  {Wyithe}}{2008}]{Geil08}
{Geil} P.~M.,  {Wyithe} J. S.~B.,  2008, \mn@doi [\mnras]
  {10.1111/j.1365-2966.2008.13159.x}, \href
  {https://ui.adsabs.harvard.edu/abs/2008MNRAS.386.1683G} {386, 1683}

\bibitem[\protect\citeauthoryear{{Geil}, {Mutch}, {Poole}, {Duffy}, {Mesinger}
  \& {Wyithe}}{{Geil} et~al.}{2017}]{Geil17}
{Geil} P.~M.,  {Mutch} S.~J.,  {Poole} G.~B.,  {Duffy} A.~R.,  {Mesinger} A.,
  {Wyithe} J. S.~B.,  2017, \mn@doi [Monthly Notices of the Royal Astronomical
  Society] {10.1093/mnras/stx1841}, \href
  {https://ui.adsabs.harvard.edu/abs/2017MNRAS.472.1324G} {472, 1324}

\bibitem[\protect\citeauthoryear{{Ghara} \& {Choudhury}}{{Ghara} \&
  {Choudhury}}{2019}]{Ghara19}
{Ghara} R.,  {Choudhury} T.~R.,  2019, arXiv e-prints, \href
  {https://ui.adsabs.harvard.edu/abs/2019arXiv190912317G} {p. arXiv:1909.12317}

\bibitem[\protect\citeauthoryear{{Ghara}, {Choudhury}, {Datta}  \&
  {Choudhuri}}{{Ghara} et~al.}{2017}]{Ghara17}
{Ghara} R.,  {Choudhury} T.~R.,  {Datta} K.~K.,   {Choudhuri} S.,  2017,
  \mn@doi [\mnras] {10.1093/mnras/stw2494}, \href
  {https://ui.adsabs.harvard.edu/abs/2017MNRAS.464.2234G} {464, 2234}

\bibitem[\protect\citeauthoryear{{Greig} \& {Mesinger}}{{Greig} \&
  {Mesinger}}{2017}]{Greig17}
{Greig} B.,  {Mesinger} A.,  2017, \mn@doi [\mnras] {10.1093/mnras/stw3026},
  \href {https://ui.adsabs.harvard.edu/abs/2017MNRAS.465.4838G} {465, 4838}

\bibitem[\protect\citeauthoryear{{Hassan}, {Liu}, {Kohn}  \& {La
  Plante}}{{Hassan} et~al.}{2019}]{Hassan19}
{Hassan} S.,  {Liu} A.,  {Kohn} S.,   {La Plante} P.,  2019, \mn@doi [\mnras]
  {10.1093/mnras/sty3282}, \href
  {https://ui.adsabs.harvard.edu/abs/2019MNRAS.483.2524H} {483, 2524}

\bibitem[\protect\citeauthoryear{{Hinshaw} et~al.,}{{Hinshaw}
  et~al.}{2013}]{Hinshaw2013}
{Hinshaw} G.,  et~al., 2013, \mn@doi [\apjs] {10.1088/0067-0049/208/2/19},
  \href {https://ui.adsabs.harvard.edu/abs/2013ApJS..208...19H} {208, 19}

\bibitem[\protect\citeauthoryear{{Hopkins}}{{Hopkins}}{2013}]{Hopkins13}
{Hopkins} P.~F.,  2013, \mn@doi [\mnras] {10.1093/mnras/sts210}, \href
  {https://ui.adsabs.harvard.edu/abs/2013MNRAS.428.2840H} {428, 2840}

\bibitem[\protect\citeauthoryear{{Katz}, {Weinberg}  \& {Hernquist}}{{Katz}
  et~al.}{1996}]{Katz96}
{Katz} N.,  {Weinberg} D.~H.,   {Hernquist} L.,  1996, \mn@doi [\apjs]
  {10.1086/192305}, \href
  {https://ui.adsabs.harvard.edu/abs/1996ApJS..105...19K} {105, 19}

\bibitem[\protect\citeauthoryear{{Kohler}, {Gnedin}, {Miralda-Escud{\'e}}  \&
  {Shaver}}{{Kohler} et~al.}{2005}]{Kohler05}
{Kohler} K.,  {Gnedin} N.~Y.,  {Miralda-Escud{\'e}} J.,   {Shaver} P.~A.,
  2005, \mn@doi [\apj] {10.1086/444370}, \href
  {https://ui.adsabs.harvard.edu/abs/2005ApJ...633..552K} {633, 552}

\bibitem[\protect\citeauthoryear{{Liu}, {Parsons}  \& {Trott}}{{Liu}
  et~al.}{2014}]{Liu2014}
{Liu} A.,  {Parsons} A.~R.,   {Trott} C.~M.,  2014, \mn@doi [\prd]
  {10.1103/PhysRevD.90.023018}, \href
  {https://ui.adsabs.harvard.edu/abs/2014PhRvD..90b3018L} {90, 023018}

\bibitem[\protect\citeauthoryear{{Ma}, {Ciardi}, {Kakiichi}, {Zaroubi}, {Zhi}
  \& {Busch}}{{Ma} et~al.}{2020}]{Ma20}
{Ma} Q.-B.,  {Ciardi} B.,  {Kakiichi} K.,  {Zaroubi} S.,  {Zhi} Q.-J.,
  {Busch} P.,  2020, \mn@doi [\apj] {10.3847/1538-4357/ab5b95}, \href
  {https://ui.adsabs.harvard.edu/abs/2020ApJ...888..112M} {888, 112}

\bibitem[\protect\citeauthoryear{{Majumdar}, {Bharadwaj}  \&
  {Choudhury}}{{Majumdar} et~al.}{2012}]{Majumdar12}
{Majumdar} S.,  {Bharadwaj} S.,   {Choudhury} T.~R.,  2012, \mn@doi [\mnras]
  {10.1111/j.1365-2966.2012.21914.x}, \href
  {https://ui.adsabs.harvard.edu/abs/2012MNRAS.426.3178M} {426, 3178}

\bibitem[\protect\citeauthoryear{{Marshall}, {Ni}, {Di Matteo}, {Wyithe},
  {Wilkins}  \& {Croft}}{{Marshall} et~al.}{2019}]{Marshall19}
{Marshall} M.~A.,  {Ni} Y.,  {Di Matteo} T.,  {Wyithe} J. S.~B.,  {Wilkins} S.,
    {Croft} R. A.~C.,  2019, arXiv e-prints, \href
  {https://ui.adsabs.harvard.edu/abs/2019arXiv191203428M} {p. arXiv:1912.03428}

\bibitem[\protect\citeauthoryear{{Mellema} et~al.,}{{Mellema}
  et~al.}{2013}]{Mellema13}
{Mellema} G.,  et~al., 2013, \mn@doi [Experimental Astronomy]
  {10.1007/s10686-013-9334-5}, \href
  {https://ui.adsabs.harvard.edu/abs/2013ExA....36..235M} {36, 235}

\bibitem[\protect\citeauthoryear{{Mesinger}, {Greig}  \& {Sobacchi}}{{Mesinger}
  et~al.}{2016}]{EOS2016}
{Mesinger} A.,  {Greig} B.,   {Sobacchi} E.,  2016, \mn@doi [\mnras]
  {10.1093/mnras/stw831}, \href
  {https://ui.adsabs.harvard.edu/abs/2016MNRAS.459.2342M} {459, 2342}

\bibitem[\protect\citeauthoryear{{Ni}, {Di Matteo}, {Gilli}, {Croft}, {Feng}
  \& {Norman}}{{Ni} et~al.}{2019}]{Ni19}
{Ni} Y.,  {Di Matteo} T.,  {Gilli} R.,  {Croft} R. A.~C.,  {Feng} Y.,
  {Norman} C.,  2019, arXiv e-prints, \href
  {https://ui.adsabs.harvard.edu/abs/2019arXiv191203780N} {p. arXiv:1912.03780}

\bibitem[\protect\citeauthoryear{{Okamoto}, {Frenk}, {Jenkins}  \&
  {Theuns}}{{Okamoto} et~al.}{2010}]{Okamoto10}
{Okamoto} T.,  {Frenk} C.~S.,  {Jenkins} A.,   {Theuns} T.,  2010, \mn@doi
  [\mnras] {10.1111/j.1365-2966.2010.16690.x}, \href
  {https://ui.adsabs.harvard.edu/abs/2010MNRAS.406..208O} {406, 208}

\bibitem[\protect\citeauthoryear{{Pober} et~al.,}{{Pober}
  et~al.}{2013}]{Pober2013}
{Pober} J.~C.,  et~al., 2013, \mn@doi [The Astronomical Journal]
  {10.1088/0004-6256/145/3/65}, \href
  {https://ui.adsabs.harvard.edu/abs/2013AJ....145...65P} {145, 65}

\bibitem[\protect\citeauthoryear{{Pober} et~al.,}{{Pober}
  et~al.}{2014}]{Pober2014}
{Pober} J.~C.,  et~al., 2014, \mn@doi [The Astrophysical Journal]
  {10.1088/0004-637X/782/2/66}, \href
  {https://ui.adsabs.harvard.edu/abs/2014ApJ...782...66P} {782, 66}

\bibitem[\protect\citeauthoryear{{Read}, {Hayfield}  \& {Agertz}}{{Read}
  et~al.}{2010}]{Read10}
{Read} J.~I.,  {Hayfield} T.,   {Agertz} O.,  2010, \mn@doi [\mnras]
  {10.1111/j.1365-2966.2010.16577.x}, \href
  {https://ui.adsabs.harvard.edu/abs/2010MNRAS.405.1513R} {405, 1513}

\bibitem[\protect\citeauthoryear{{Spergel} et~al.,}{{Spergel}
  et~al.}{2015}]{Spergel15}
{Spergel} D.,  et~al., 2015, arXiv e-prints, \href
  {https://ui.adsabs.harvard.edu/abs/2015arXiv150303757S} {p. arXiv:1503.03757}

\bibitem[\protect\citeauthoryear{{Springel} \& {Hernquist}}{{Springel} \&
  {Hernquist}}{2003}]{Springel03}
{Springel} V.,  {Hernquist} L.,  2003, \mn@doi [\mnras]
  {10.1046/j.1365-8711.2003.06206.x}, \href
  {https://ui.adsabs.harvard.edu/abs/2003MNRAS.339..289S} {339, 289}

\bibitem[\protect\citeauthoryear{{Vogelsberger}, {Genel}, {Sijacki}, {Torrey},
  {Springel}  \& {Hernquist}}{{Vogelsberger} et~al.}{2013}]{Vogelsberger13}
{Vogelsberger} M.,  {Genel} S.,  {Sijacki} D.,  {Torrey} P.,  {Springel} V.,
  {Hernquist} L.,  2013, \mn@doi [\mnras] {10.1093/mnras/stt1789}, \href
  {https://ui.adsabs.harvard.edu/abs/2013MNRAS.436.3031V} {436, 3031}

\bibitem[\protect\citeauthoryear{{Vogelsberger} et~al.,}{{Vogelsberger}
  et~al.}{2014a}]{Vogelsberger14}
{Vogelsberger} M.,  et~al., 2014a, \mn@doi [\mnras] {10.1093/mnras/stu1536},
  \href {https://ui.adsabs.harvard.edu/abs/2014MNRAS.444.1518V} {444, 1518}

\bibitem[\protect\citeauthoryear{{Vogelsberger} et~al.,}{{Vogelsberger}
  et~al.}{2014b}]{Krumholz11}
{Vogelsberger} M.,  et~al., 2014b, \mn@doi [\mnras] {10.1093/mnras/stu1536},
  \href {https://ui.adsabs.harvard.edu/abs/2014MNRAS.444.1518V} {444, 1518}

\bibitem[\protect\citeauthoryear{{Waters}, {Wilkins}, {Di Matteo}, {Feng},
  {Croft}  \& {Nagai}}{{Waters} et~al.}{2016}]{Waters16}
{Waters} D.,  {Wilkins} S.~M.,  {Di Matteo} T.,  {Feng} Y.,  {Croft} R.,
  {Nagai} D.,  2016, \mn@doi [\mnras] {10.1093/mnrasl/slw100}, \href
  {https://ui.adsabs.harvard.edu/abs/2016MNRAS.461L..51W} {461, L51}

\bibitem[\protect\citeauthoryear{{Wilkins}, {Feng}, {Di Matteo}, {Croft},
  {Lovell}  \& {Waters}}{{Wilkins} et~al.}{2017}]{Wilkins17}
{Wilkins} S.~M.,  {Feng} Y.,  {Di Matteo} T.,  {Croft} R.,  {Lovell} C.~C.,
  {Waters} D.,  2017, \mn@doi [\mnras] {10.1093/mnras/stx841}, \href
  {https://ui.adsabs.harvard.edu/abs/2017MNRAS.469.2517W} {469, 2517}

\bibitem[\protect\citeauthoryear{{Wyithe}, {Loeb}  \& {Barnes}}{{Wyithe}
  et~al.}{2005}]{Wyithe05}
{Wyithe} J. S.~B.,  {Loeb} A.,   {Barnes} D.~G.,  2005, \mn@doi [\apj]
  {10.1086/497160}, \href
  {https://ui.adsabs.harvard.edu/abs/2005ApJ...634..715W} {634, 715}

\makeatother
\end{thebibliography}



\appendix
\section{fit systematics}\label{sec:rand}
\begin{figure}
\includegraphics[width=\linewidth]{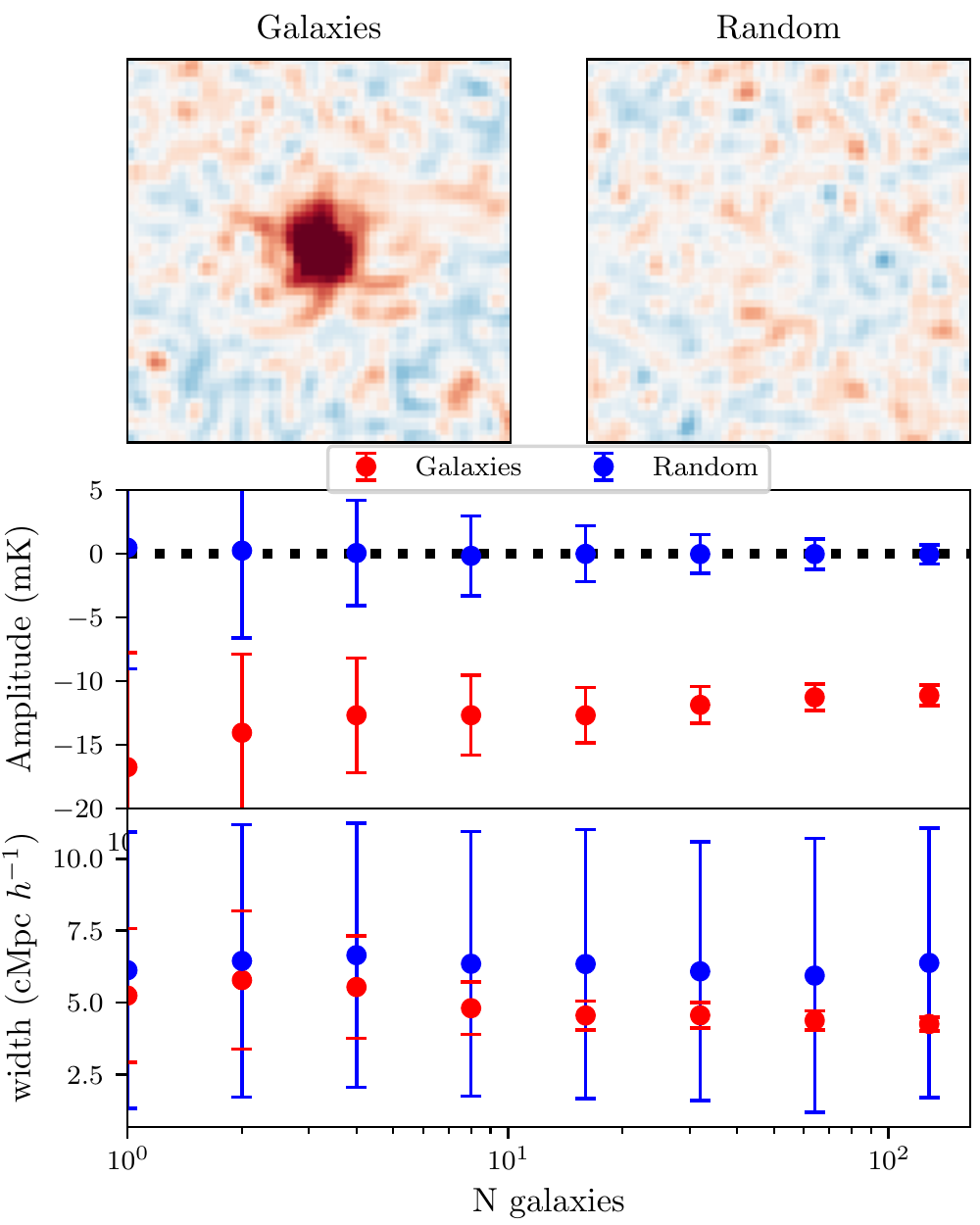}
\caption{Fits around stacks of up to 128 bright galaxies and random pointings compared. Top panels: $40 \times 40 \times 5 \rm{h^{-1}cMpc}$ slices of 30 stacked galaxy (left) and random (right) pointings. Middle: Gaussian fit amplitude for both galaxy and random pointings. Bottom: Gaussian fit widths for both cases. Random pointings are fit closely to zero amplitude, and show a wide range of widths, consistent with noise.}
\label{fig:random}
\end{figure}

In this appendix we first test the systematics of our fitting process, and to justify our usage of the Gaussian fit amplitude as our signal-to-noise metric. We apply the same sampling and fitting method to random positions in the \textsc{Bluetides} volume, to examine the case of non-detection. Firgure \ref{fig:random} shows that random pointings are fit very closely to zero amplitude, and shows a very high variance in width, since density fluctuations and noise can appear on different scales near the center of the image. Since these widths include those of the real HII regions, we only use the height of the profiles to measure signal-to-noise.

\begin{figure}
\includegraphics[width=\linewidth]{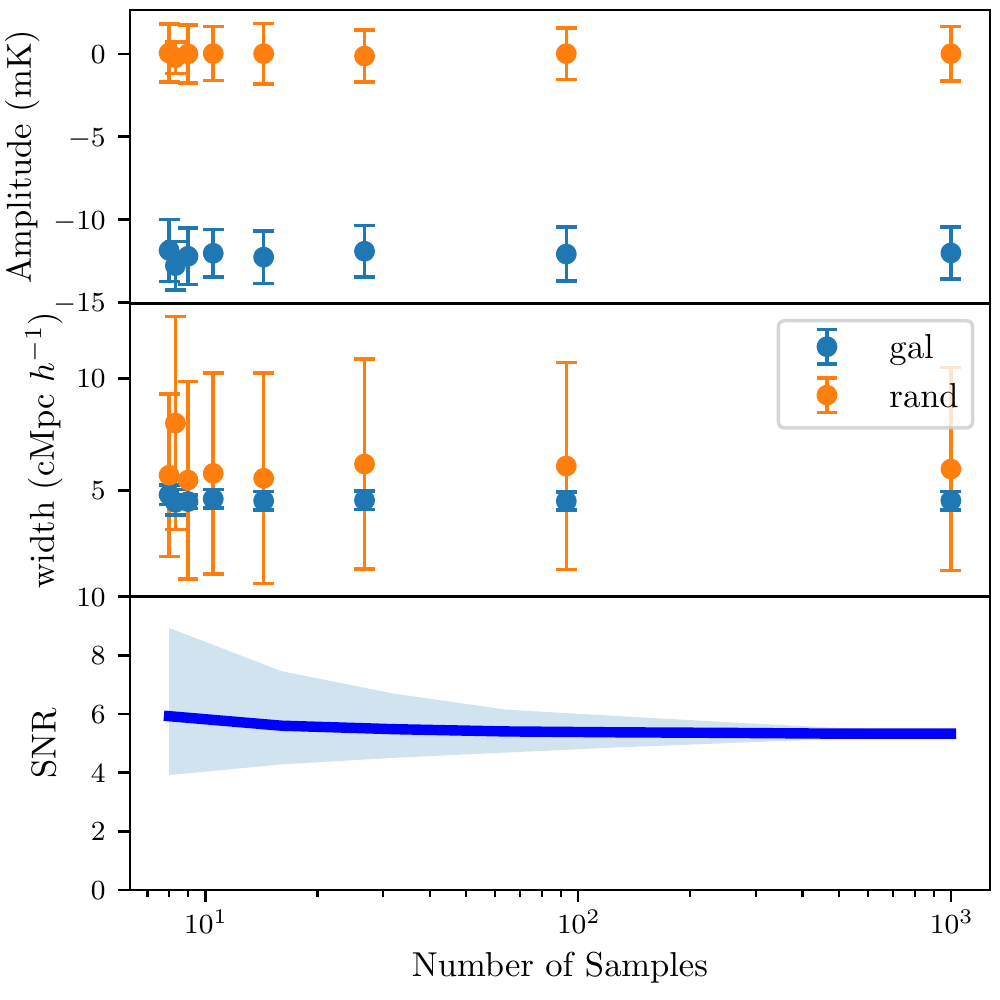}
\caption{Fit parameters for different numbers of galaxy samples. Top: Gaussian fit amplitudes. Middle: Gaussian fit widths. Bottom: signal to noise ratios, where the shaded area represents the 95th percentile limits of SNR when the process is repeated many times. As more galaxy samples are taken, variation in parameters and SNR decreases. However, variation in SNR remains under 10\% for 50 - 1000 galaxy samples}
\label{fig:mcloops}
\end{figure}

We have also tested the convergence of the model for different numbers of bootstrap samples of 30 galaxies. Figure \ref{fig:mcloops} shows that as more galaxy samples are taken, variation in both the parameters and signal to noise ratio decreases. However, both the parameters and SNR vary by less than 10\% for 100 - 1000 galaxy samples.

\begin{figure}
\includegraphics[width=\linewidth]{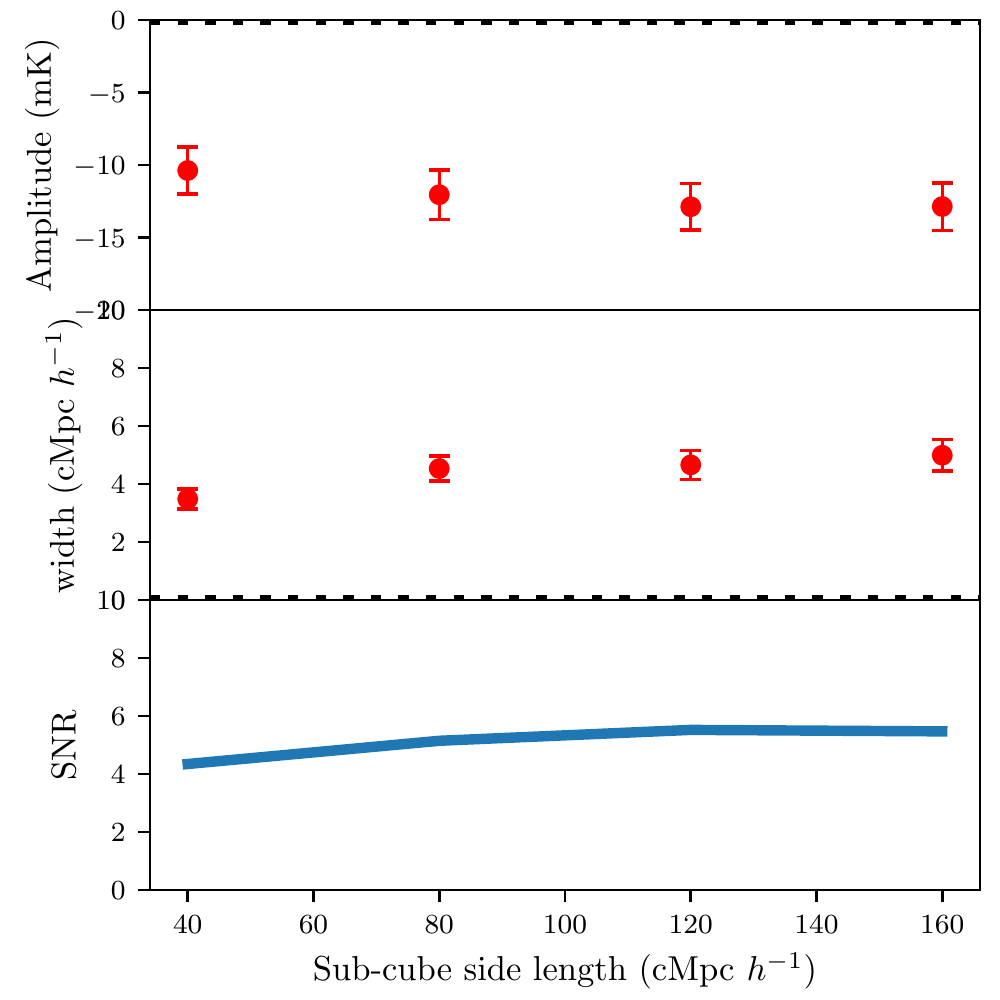}
\caption{Fit parameters from different size sub-cubes. Top: Gaussian fit amplitudes. Middle: Gaussian fit widths. Bottom: Signal to noise ratio. Results for sub-cubes larger than $80^3$ cells shows little change in both fit parameters and SNR}
\label{fig:boxsize}
\end{figure}

The size of the sub-cubes that we stack has implications for the Fourier-space resolution when we add thermal noise and foregrounds. To test how sub-cube size affects the gaussian parameters, we compare stacks of sub-cubes  of side length 40, 80, 120 and 160 $\rm{h^{-1}cMpc}$. We show in Figure \ref{fig:boxsize} that above 80 $\rm{h^{-1}cMpc}$, differences in parameters and SNR remain under 10 percent.

\bsp	
\label{lastpage}
\end{document}